\documentclass[pdflatex]{sn-jnl}

\usepackage{tabularx}
\usepackage{xcolor}
\usepackage{xpatch}
\usepackage{bm}
\usepackage{hyperref, url}
\usepackage[super]{nth}
\usepackage{adjustbox}
\usepackage{placeins}
\usepackage{rotating}
\usepackage{tabularx}
\usepackage{dsfont}
\usepackage{booktabs}
\usepackage{longtable}
\usepackage{array}
\usepackage{multirow}
\usepackage{wrapfig}
\usepackage{float}
\usepackage{colortbl}
\usepackage{pdflscape}
\usepackage{tabu}
\usepackage{threeparttable}
\usepackage{threeparttablex}
\usepackage[normalem]{ulem}
\usepackage{makecell}
\usepackage{xcolor}
\usepackage{xpatch}
\usepackage{hyperref}
\usepackage{multirow}
\usepackage{subcaption}
\usepackage{enumitem}
\usepackage{pdfpages}
\usepackage{eurosym}
\usepackage{etoolbox}
\usepackage{nicematrix}
\usepackage{subcaption}
\usepackage{graphicx}
\usepackage{doi}
\usepackage{csquotes}
\usepackage{siunitx} 
\usepackage{graphicx}%
\usepackage{multirow}%
\usepackage{amsmath,amssymb,amsfonts}%
\usepackage{amsthm}%
\usepackage{mathrsfs}%
\usepackage[title]{appendix}%
\usepackage{xcolor}%
\usepackage{textcomp}%
\usepackage{manyfoot}%
\usepackage{booktabs}%
\usepackage{algorithm}%
\usepackage{algorithmicx}%
\usepackage{algpseudocode}%
\usepackage{listings}%

\usepackage[backend=biber,
            style=authoryear,
            sorting=nyt,    
            maxbibnames=99,  
            doi=true,       
            maxcitenames=2,     
            mincitenames=1,     
            url=false,        
            dashed=false,
            isbn=false]{biblatex}

\AtEveryBibitem{\clearfield{month}} 

\AtBeginBibliography{\small}

\newcommand{\theTitle}{A spatio-temporal statistical model for property valuation at country-scale with adjustments for regional submarkets}

\makeatletter
\xpatchcmd\NAT@citex
 {%
  \@citea\NAT@hyper@{%
    \NAT@nmfmt{\NAT@nm}%
    \hyper@natlinkbreak{\NAT@aysep\NAT@spacechar}{\@citeb\@extra@b@citeb}%
    \NAT@date
  }%
 }
 {%
  \@citea
  \NAT@nmfmt{\NAT@nm}%
  \NAT@aysep\NAT@spacechar
  \NAT@hyper@{\NAT@date}%
 }
 {}{}
\xpatchcmd\NAT@citex
 {%
  \@citea\NAT@hyper@{%
    \NAT@nmfmt{\NAT@nm}%
    \hyper@natlinkbreak{\NAT@spacechar\NAT@@open\if*#1*\else#1\NAT@spacechar\fi}%
    {\@citeb\@extra@b@citeb}%
    \NAT@date
  }%
 }
 {
  \@citea
    \NAT@nmfmt{\NAT@nm}%
    \NAT@spacechar\NAT@@open\if*#1*\else#1\NAT@spacechar\fi
    \NAT@hyper@{\NAT@date}%
 }
{}{}
\makeatother

\DeclareCaptionFormat{custom}
{%
    \small\textbf{#1}\quad\small #3
}
\makeatletter
\renewcommand{\fnum@figure}{Fig \thefigure}
\makeatother

\usepackage{setspace}
\usepackage{parskip}
\usepackage{amsmath}
\usepackage{amsfonts}
\usepackage{amssymb}

\usepackage{graphicx}
\graphicspath{{./}{./Figures/}}

\usepackage[english]{babel}

\usepackage{hhline} 
\usepackage{graphicx} 

\newcommand{\rsq}{$R^2$ }

\usepackage{hyperref}

\hypersetup{
unicode=true,                 
pdftoolbar=true,              
pdfmenubar=true,              
pdffitwindow=true,            
pdftitle={\theTitle},         
pdfauthor={},  
pdfsubject={},                
pdfnewwindow=true,            
pdfkeywords={},               
colorlinks=true,              
linkcolor=blue,                
citecolor=blue,                
filecolor=blue,               
urlcolor=blue                
}

\usepackage{fontawesome5}

\makeatletter
\renewcommand{\@maketitle}{%
    \newpage\null%
    \if@remarkboxon
        \vbox to 0pt{\vspace*{-78pt}\hspace*{-18pt}\FMremark}%
    \else
        \vskip21pt%
    \fi
    \hsize\textwidth\parindent0pt%
    {\hbox to \textwidth{{\Artcatfont\ArtType\hfill}\par}}%
    \ifx\@title\empty\else%
        \removelastskip\vskip20pt\nointerlineskip%
        {\Titlefont\@title\par}%
    \fi%
    \ifx\@subtitle\empty\else%
        \vskip9pt%
        {{\SubTitlefont\@subtitle\par}}%
    \fi%
    \ifnum\aucount>0
        \global\punctcount\aucount%
        \vskip20pt%
        \artauthors\par%
        {\vskip7pt
         \addressfont\auaddress\par%
         \removelastskip\vskip24pt%
         \ifnum\emailcnt>0\relax%
             \ifx\corrauthemail\@empty\else{\ifnum\aucount>1*\fi}%
             Corresponding author(s). E-mail(s): \corrauthemail\par\fi%
             \ifx\authemail\@empty\else Contributing authors: \authemail\fi%
         \fi%
         \ifequalcont{\par$^{\dagger}$\@equalconttext\par}\fi%
         \removelastskip\vskip24pt%
         \ifpresentaddress{\par\@presentaddresstext\par}\fi%
        }%
    \fi%
    {\printabstract\par}%
    {\printkeywords\par}%
    \ifx\@pacs\empty\else%
        \loop\ifnum\PacsCount>0%
            \csname\romannumeral\PacsTmpCnt StorePacsTxt\endcsname\par%
            \StepDownCounter{\PacsCount}%
            \StepUpCounter{\PacsTmpCnt}%
        \repeat%
    \fi%
    \vfill
    \ifx\@thanks\@empty\else%
        \begin{center}
            \@thanks
        \end{center}%
    \fi%
    \removelastskip\vskip36pt\vskip0pt%
}
\makeatother
\makeatletter
\def\thanks#1{\protected@xdef\@thanks{\@thanks
        \protect\footnotetext{#1}}}
\makeatother

\begin{document}

\title[Article Title]{\theTitle}

\author[1]{\fnm{Brian} \sur{O'Donovan}%
  \thanks{$^{\dagger}$ Present address: School of Engineering, IDCOM, University of Edinburgh, Edinburgh, UK.}%
  \thanks{\faEnvelope~Corresponding author: \href{mailto:bodonov@ed.ac.uk}{b.odonov@ed.ac.uk}}$^{\dagger}$
}
\author[2,3]{\fnm{Andrew} \sur{Finley}}
\author[1]{\fnm{James} \sur{Sweeney}}

\affil[1]{\orgdiv{Department of Mathematics and Statistics}, \orgname{University of Limerick}, \orgaddress{\city{Limerick}, \postcode{V94 T9PX}, \country{Ireland}}}

\affil[2]{\orgdiv{Department of Forestry}, \orgname{Michigan State University},
\orgaddress{\city{East Lansing}, \postcode{MI}, \country{USA}}}

\affil[3]{\orgdiv{Department of Statistics and Probability}, \orgname{Michigan State University},
\orgaddress{\city{East Lansing}, \postcode{MI}, \country{USA}}}

\abstract{Valuing residential property is inherently complex, requiring consideration of numerous environmental, economic, and property-specific factors. These complexities present significant challenges for automated valuation models (AVMs), which are increasingly used to provide objective assessments for property taxation and mortgage financing. The challenge of obtaining accurate and objective valuations for properties at a country level, and not just within major cities, is further compounded by the presence of multiple localised submarkets—spanning urban, suburban, and rural contexts—where property features contribute differently to value. Existing AVMs often struggle in such settings: traditional hedonic regression models lack the flexibility to capture spatial variation, while advanced machine learning approaches demand extensive datasets that are rarely available. In this article, we address these limitations by developing a robust statistical framework for property valuation in the Irish housing market. We segment the country into six submarkets encompassing cities, large towns, and rural areas, and employ a generalized additive model that captures non-linear effects of property characteristics while allowing feature contributions to vary across submarkets. Our approach outperforms both machine learning–based and traditional hedonic regression models, particularly in data-sparse regions. In out-of-sample validation, our model achieves \rsq values of 0.70, 0.84, and 0.83 for rural areas, towns, and Dublin, respectively, compared to 0.52, 0.71, and 0.82 from a random forest benchmark. Furthermore, the temporal dynamics of our model align closely with reported inflation figures for the study period, providing additional validation of its accuracy.}




\maketitle

\newpage

\section{Introduction}

Accurate property valuation is critical due to the significant financial investment involved in home purchases and their role in mortgage financing, taxation, and property indices \parencite{Eurostat2013}. Commercial and public interest drive the demand for precision in automated valuation models (AVM). Traditional estimates often rely on outdated methods, while the more favoured statistical approaches use the properties' unique structural, neighbourhood, and locational characteristics to predict property value.


In the Republic of Ireland, housing related research has focused either at national scale or on specific urban environments such as the capital city of Dublin.  A national house price index for Ireland was developed by \textcite{o2011constructing} using data from the Residential Property Price Index (RPPI) and financial institutions. O'Hanlon outlines a lack of postcode regions at a national level, which complicates the development of a national mixed adjustment index. The urban areas of Cork, Galway and Limerick are analysed without any micro-locational records thus limiting the granularity of results. To overcome the lack of detailed sales records of Irish dwellings, \textcite{Maguire2016} developed a property price index using sparse national data, namely, house location, sale date, and price. The current Central Statistics Office (CSO) residential property price index uses data from multiple government bodies to train a hedonic model of four independent variables, namely: property living area, property type (terraced/detached/semi-detached), Eircode routing key, and an area deprivation index. Observations are segregated into 13 subareas based on geographical location, and 13 independent regression models are trained within each subarea \parencite{cso_rppi2016}. \textcite{mcquinn2024residential} examine the land value of properties in Cork, Dublin, Galway and Limerick using property sale transaction data from 2010 to 2017, where land value is estimated as the difference between property price and the cost of rebuilding the property. The authors note greater fluctuations in land prices compared to property prices during the period, with Dublin demonstrating the most volatility in prices. In Dublin, \textcite{mayor2009hedonic} studies the influence of green spaces on prices, and \textcite{roche2001rise} models property prices using economic factors. \textcite{Kitchin2013} identifies issues with the property sales records in Ireland, which are inadequate in spatial terms and lack essential features.


The traditional modelling approach for property prices is the hedonic multiple linear regression model \parencite{Rosen1974}. This additive approach models the value of a property as the intrinsic value of its attributes and remains the recommended approach by \textcite{Eurostat2013} for government departments. This involves fitting a unitary, global equation to the entire market, which often proves too inflexible to accommodate local and regional relationships between the response and predictors, resulting in several shortcomings, including limited spatial representativeness, vulnerability to omitted-variable bias, and poor performance in capturing non-linear or structured patterns within the covariates \parencite{Hurley2022}.  The underlying issues common to spatial data are spatial autocorrelation, such that neighbouring data points share similar values of the response variable, and spatial heterogeneity, whereby the relationships between attributes vary over space, leading to structural instability under the assumption of stationary relationships \parencite{anselin1988spatial}. 



Generalized additive models (GAMs) extend linear regression by estimating the response as a sum of nonparametric functions of the predictors, typically via smoothing splines \parencite[pp.~289-311]{james2023}. This removes the need to specify transformations or functional forms \textit{a priori}, and each spline can be visualised for interpretability. GAMs have been shown to outperform parametric and polynomial models in property valuation, particularly with comprehensive datasets \parencite{pace1998appraisal, gelfand1998spatio, panduro2013classification}, although \textcite{shimizu2014nonlinearity} report limited gains in out-of-sample prediction for some cases. In studies of Dublin house prices, GAMs capture spatial variation effectively \parencite{dupre2020urban, Hurley2022}. Hurley and Sweeney model a closed subset of data using smoothing splines, while Dupré highlights the sensitivity of continuous Bayesian splines to geographical features such as rivers or coastal boundaries. \textit{Finite area smoothing}, including soap film smoothing, provides a computationally efficient approach to address these boundary effects \parencite{wood2008soap}.


Market segmentation, where properties are grouped into submarkets with similar characteristics, improves valuation accuracy \parencite{Basu1998, bourassa1999defining, AllenC.Goodman2003}. There is extensive literature on the methods of assigning houses to submarkets. Common methods include the use of: demographic data \parencite{bourassa2010predicting}; postcode regions \parencite{AllenC.Goodman2003, Hurley2022}; census blocks and school districts \parencite{Basu1998}. GAMs provide a data-driven way to model such segmentation by allowing relationships between property prices and their determinants to vary smoothly across space. This enables the approximation of submarket structures without predefined boundaries. Local relationships are thus applied within the GAM framework, rather than imposing global relationships to local phenomena. \textcite{dearmon2024local} identify and model local ``comps" of comparable properties based on realtor knowledge. A spatially-varying coefficient model is applied by \textcite{comber2024multiscale} using GP splines parametrised at observation locations.


Machine learning (ML) is widely used in automated property valuation, with tools like Zillow's \textit{Zestimate} being popular in the USA. Unlike statistical techniques, ML makes few assumptions and performs well with large datasets \parencite{das2021boosting}. Even on smaller datasets, ensemble methods such as boosting can outperform Gaussian process regression \parencite{lahmiri2023comparative}. Comparisons in Ireland show that \textit{k}-Nearest Neighbours approach struggles with sparse data, while random forests (RF) predict means more accurately than GAMs but less reliably for prediction intervals \parencite{Hurley2022}. Data quality issues, including limited spatial detail, remain a challenge in Irish property studies \parencite{o2011constructing, Maguire2016, rabiei2021gap}.

In this article, we develop a national submarket model (S-GAM) for Ireland that integrates Gaussian process (GP) splines and Markov random fields (MRF) to capture spatial contributions. We compare the S-GAM to a conventional hedonic regression and a national-level GAM (N-GAM), and a RF using 29,458 property listings from January to December 2022. The article is structured as follows: Section \ref{sec:background} introduces the market and data; Section \ref{sec:Methodology} describes the models; results are presented in Section \ref{sec:results}; and Section \ref{sec:discussion} concludes with main findings and future directions.

\section{Background}\label{sec:background}

\subsection{The Property Market in Ireland}

There are 2.12 million habitable dwellings in Ireland \parencite{cso_stock2022}, and Ireland has the highest share of people living in houses among all European Union (EU) member states, with a proportion of 90\% compared to the EU average of 53\% \parencite{eurostat2021}.  The average Irish dwelling is transacted every 60 years \parencite{Maguire2016}, and the infrequency of property transactions has been a key concern for traditional valuation methods \parencite{Hurley2022, o2011constructing}. 
In 2022, The \textcite{cso_rppi2022} reported 50,025 dwelling purchases at market value were submitted to the Revenue Commissioners, the Irish governmental taxation organisation, with a median sale price of \euro305,000.

Ireland comprises 26 counties, including Limerick, Cork, Galway, and County Dublin, which accounts for approximately 25\% of the Irish housing stock \parencite{cso_stock2022}. The Eircode is the national postcode system in Ireland\footnote{The Eircode system was developed in 2015 by Capita Business Support Services Ireland using a database supplied by An Post GeoDirectory DAC (a subsidiary of An Post) \hyperlink{https://www.eircode.ie/what-is-eircode}{Eircode.ie}.}, the unique address-based code contains seven characters, the first three of which are a routing key referring to a specific area, for example ``V94'' corresponds to Limerick. Overall, 139 routing keys correspond to areas of varying size, with 15,000 addresses in each, on average. The remaining characters of the Eircode are randomly organised property codes, so adjacent properties have completely different Eircodes. In a geographic classification of Irish areas, \textcite{brunsdon2018open} outlines the presence of complex pricing structures across Eircode routing key areas (henceforth known as Eircode regions) and the potential of such areas for market analysis.

\subsection{The Irish National Property Price Dataset}

The Irish national property price dataset has details of over 40,000 property transactions across Ireland ranging from January 2022 to December 2022, inclusive. The data was provided by 4Property Ltd with the sales price obtained from the Residential Property Price Register from the 
Property Services Regulatory Authority (PSRA).

Listings missing values for price or location coordinates are excluded, resulting in 29,458 observations. Property characteristics are explored across the 26 counties in the Republic of Ireland and the 139 unique Eircode regions. The variables of interest are described in Table \ref{tab:variables}, and the data cleaning methods are outlined in the following section.

\begin{table}[t!]
\centering
\footnotesize
\caption{\label{tab:variables}Variable names and descriptions}
\begin{tabular}{@{}%
    >{\raggedright\arraybackslash}p{0.17\linewidth}%
    >{\raggedright\arraybackslash}p{0.8\linewidth}@{}}
\toprule
Variable & Description\\
\midrule
Price & Property sale price (\euro)\\
\addlinespace
Month & Month of property sale\\
\addlinespace
Area & Area of the property from Cork, Dublin, Galway, Limerick, Towns or Rural\\
\addlinespace
Longitude & Longitude coordinate of observation (World Geodetic System WGS84)\\
\addlinespace
Latitude & Latitude coordinate of observation (World Geodetic System WGS84)\\
\addlinespace
Eircode & Eircode Routing Key of the property\\
\addlinespace
Baths & The number of bathrooms in the property\\
\addlinespace
Beds & The number of bedrooms in the property\\
\addlinespace
Size & Internal area of the property in square metres (\unit{\metre\squared})\\
\addlinespace
Property Type & Property type categories from detached, semi-detached, townhouse, terraced, end-of-terrace, apartment, duplex\\
\addlinespace
BER & Building energy rating of the property on a scale from A to G\\
\addlinespace
Description & Free text property description from the listing website\\
\bottomrule
\end{tabular}
\end{table}

\subsubsection{Data Cleaning and Model Covariates}

The majority of property listings contain information on the property type, the number of bedrooms and the number of bathrooms, in addition to a text description which is mined for additional structural and design characteristics of the property. Properties additionally contain a building energy rating (BER)\footnote{The BER is assigned by the Sustainable Energy Authority of Ireland \hyperlink{https://www.seai.ie/home-energy/building-energy-rating-ber/}{SEAI.ie}. The rating is on a scale from A (high energy efficiency) to G (low energy efficiency).}.
All properties have a raw text address, sale price, and locational coordinates (longitude and latitude). 
Entries with missing values for property characteristics and no text description were removed from the data as no information could be retrieved. 

Property text descriptions are text-mined to assign a \textit{property type} to each entry. Commercial properties were excluded from this analysis. Taking a similar approach to \textcite{adair1996hedonic}, distinct levels outlining the structural characteristics of houses were created. These categorise houses as detached, semi-detached, terrace, end-of-terrace or townhouse. Each house is assigned to one of these property types based on the text-mined keywords. A similar approach is repeated for the apartment category, where keywords include ``studio'', ``apartment'', ``penthouse'' and ``flat''. Duplex properties remain a unique property type. 

Missing values for the number of bedrooms and bathrooms are imputed using text mining to extract the number of such rooms or the count of key phrases. Similarly, missing values for \textit{size} are imputed by extracting numerical values from the text description, or by combining the individually states room sizes. A string search is used to input the missing values for the BER. Property features are identified using string searches of key phrases based on similar words or accounting for spelling errors. The features identified and corresponding descriptions are outlined in Table \ref{tab:dummy_variables}.

\begin{table}[t!]
\centering
\caption{\label{tab:dummy_variables}Dummy variables and descriptions}
\begin{tabular}{@{}%
    >{\raggedright\arraybackslash}p{0.7\linewidth}%
    >{\raggedleft\arraybackslash}p{0.25\linewidth}@{}}
\toprule
Description & Count\\
\midrule
Attic Conversion & 707\\
Garden & 19,686\\
Cul-de-sac & 4,490\\
Garage & 9,304\\
Renovated Property & 4,137\\
Period Property & 619\\
South Facing Property & 3,801\\
Ground Floor Apartment & 1,141\\
Second Floor Apartment & 382\\
Penthouse Apartment & 147\\
New Property & 6,673\\
\bottomrule
\end{tabular}
\end{table}

Erroneous coordinate assignments are corrected using a \textit{Google Maps} API, and Cartesian coordinates are derived from the longitude and latitude values of each observation using a Pseudo Mercator projection, a cylindrical map projection for 85.06°S and 85.06°N \parencite{stefanakis2017web}. The \textit{sp} package developed by \textcite{sp_package} is used to visualise and analyse the spatial data with polygons. Since not all property addresses list an Eircode, we assign Eircode regions to observations by merging a shapefile of Irish Eircode regions with the data.

\begin{figure}[!t]
  \centering
  \begin{minipage}[b]{0.58\textwidth}
        \includegraphics[trim={5cm 0 0 0}, width = 1.25\linewidth]{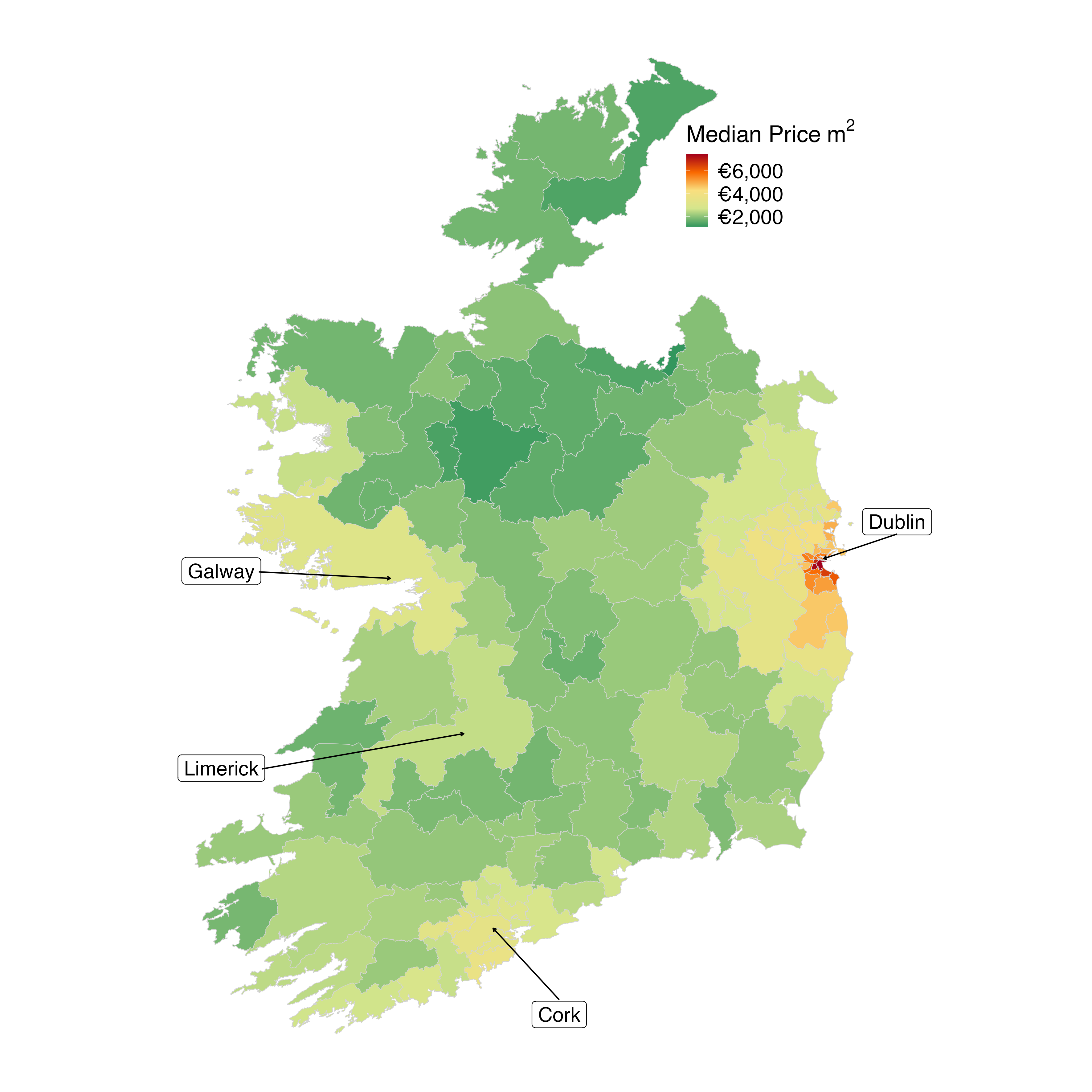}
        \caption{\label{fig:map_median_ecode}
    \small  Median values of Eircode regions with the approximate location of cities highlighted with arrows}  
  \vspace{2cm}
  \end{minipage}
  \hfill
  \begin{minipage}[b]{0.4\textwidth}
        \includegraphics[trim={4cm 0 0 0}, width = 1.5\linewidth]{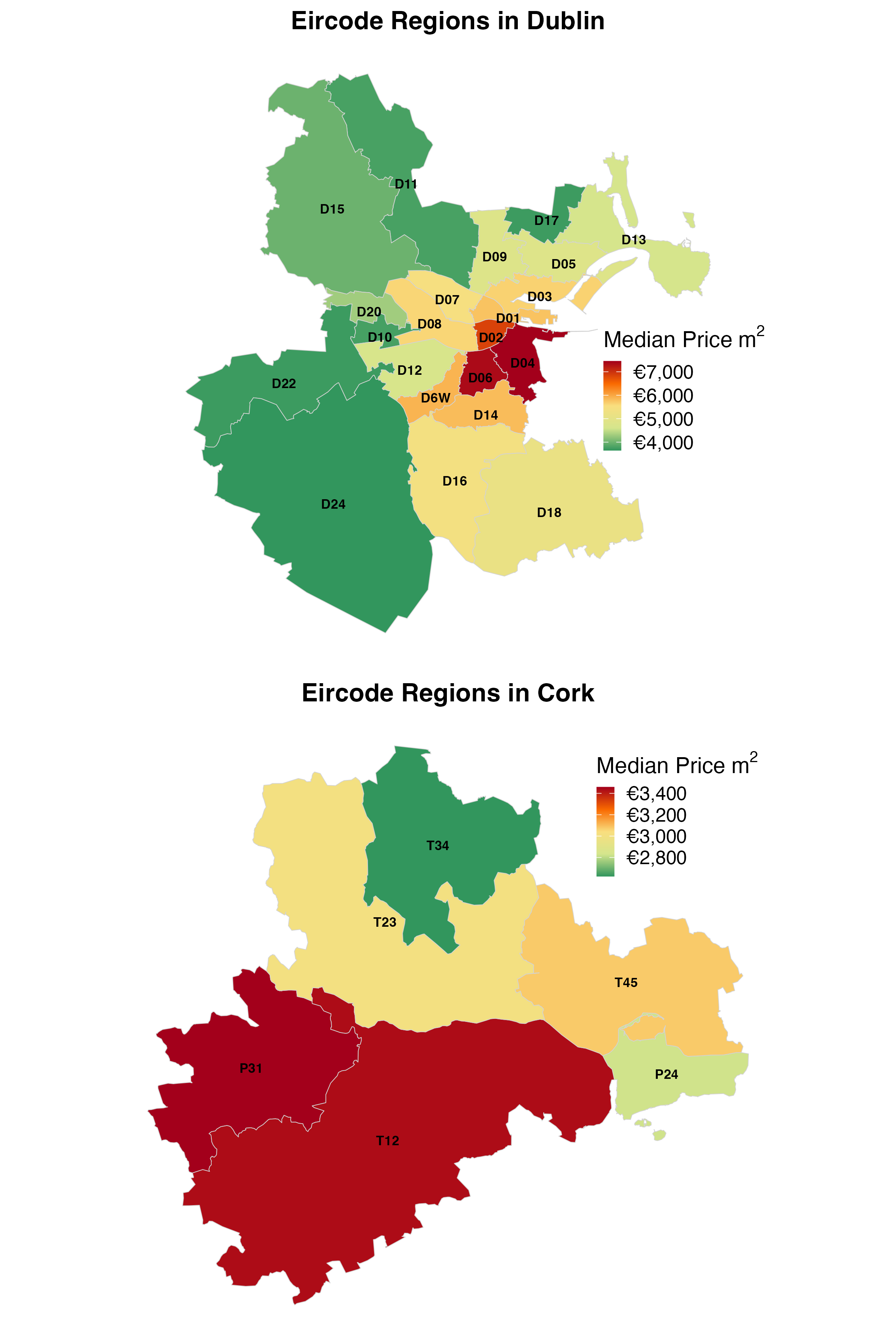}
    \caption{\label{fig:cork_dub_eircode_maps}
    \small   Median values of Eircode regions surrounding Dublin city and Cork city with individual scales}
  \end{minipage}
\end{figure}

A map of the Eircode regions in Ireland is shown in Figure \ref{fig:map_median_ecode}. Regions are coloured by the median price per $m^2$ for corresponding properties. The locations of the cities of Dublin, Galway, Cork and Limerick are highlighted with arrows. The median property value in Eircode regions around Dublin is the highest, while Eircode regions in the midlands and north have the lowest median values.

\newpage

\newpage

\subsubsection{Defining Submarkets}

\begin{figure}[h!]
  \centering
  \begin{minipage}[t]{0.38\textwidth}
    \centering
    \vspace{-7cm} 
    \footnotesize
    \captionof{table}{Submarket descriptions with the number of Eircode regions}
    \label{tab:area_desc}
    \begin{tabular}{l p{2.3cm} p{1cm}}
      \toprule
      Submarket & Description & Eircode Regions\\
      \midrule
      Cork & Cork City & 4\\
      Dublin & Dublin City & 28\\
      Galway & Galway City & 1\\
      Limerick & Limerick City & 1\\
      Rural & Areas not in towns or cities &70\\
      Towns & Towns with over 20,000 inhabitants & 35\\
      \textbf{Total} &  & \textbf{139}\\
      \bottomrule
    \end{tabular}

  \end{minipage}
  \hfill
  \begin{minipage}[t]{0.55\textwidth}
    \centering
    \includegraphics[trim={4cm 0 0 0},clip,width=1\textwidth]{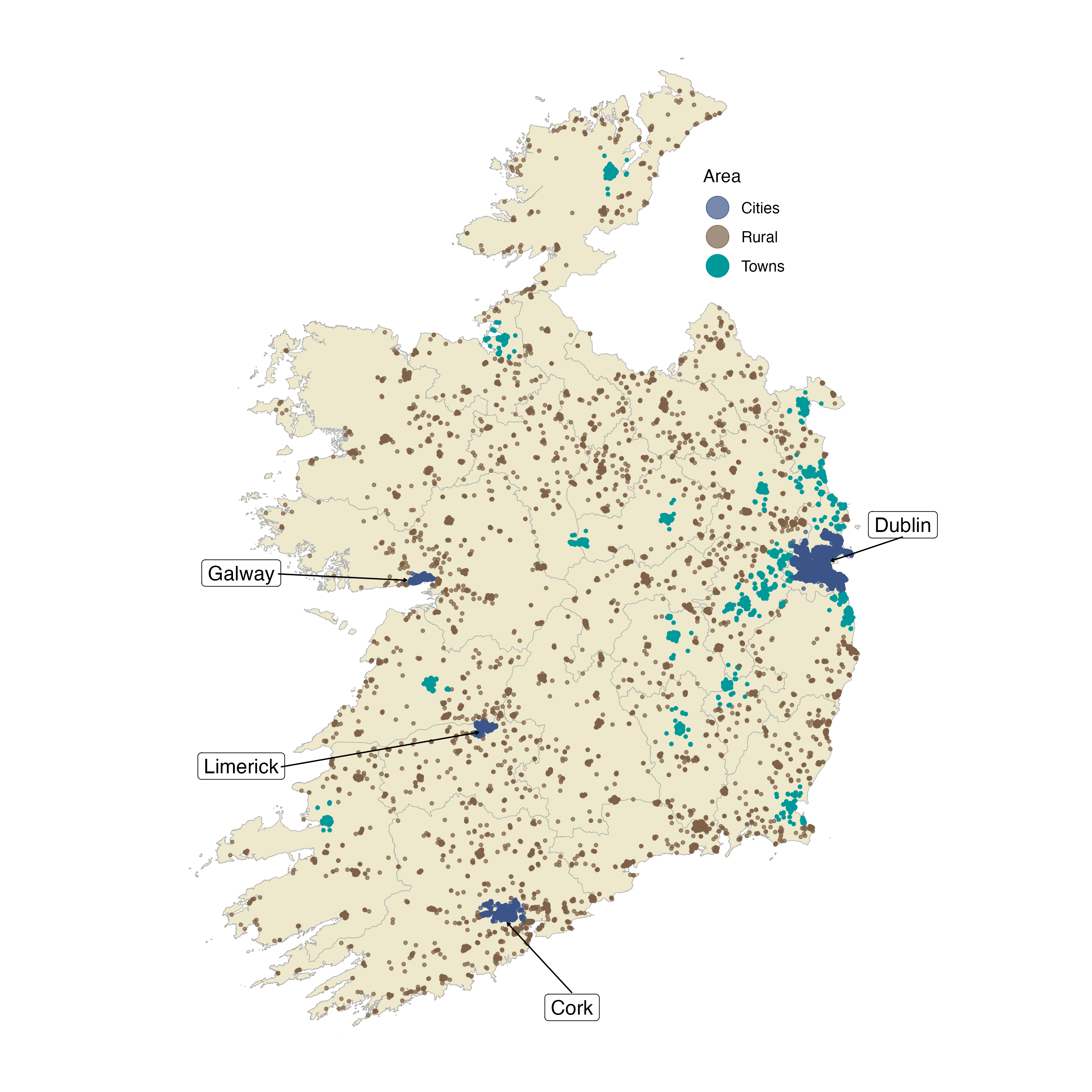}
    \caption{\small Map of observations coloured by submarket}
    \label{fig:subarea_map_lbls}
  \end{minipage}
\end{figure}

In this study, the Irish property market is divided into six distinct submarkets: the cities of Cork, Dublin, Galway, and Limerick; a group of large towns; and the remaining rural properties. Observations are assigned to the cities by overlaying shapefiles defined by each of the city boundaries. We filtered 20 towns from the \textcite{cso_towns2022} with at least 20,000 inhabitants, and selected properties within 10km of each town centre. The presence of geographical clusters in Ireland has been explored by \textcite{brunsdon2018open}, who identified areas consisting of rural communities, mature suburbs, and ``commuterland'', among others. The submarkets are described in Table \ref{tab:area_desc} and the corresponding property characteristics within submarkets are listed in Table \ref{tab:subarea_stats}

The observations are plotted on a map of Ireland in Figure \ref{fig:subarea_map_lbls} coloured by their submarket specification. The areas of Cork, Dublin, Galway and Limerick are grouped and coloured as cities to aid visualisation. There is a high density of observations in Dublin and the surrounding towns. In comparison, properties in the midlands, north and west of the country are sparsely distributed.

\newpage

\newpage
\begin{landscape}
\begin{table}[!htbp]
\centering
\caption{\label{tab:subarea_stats}Summary of property types, property characteristics and price per $m^2$ in submarkets}
\vspace{3cm}
\begin{tabular}{
p{2.8cm}p{1.25cm}p{1.25cm}p{1.25cm}p{1.25cm}p{1.25cm}p{1.25cm}p{1.25cm}
p{1.3cm} 
p{1.3cm} 
p{1.5cm}
}
\toprule
Submarket & Detached House & Semi-detached House & Terraced House & End-of-terrace House & Townhouse & Apartment & Duplex Property & Median Price \!\!per $m^2$ & Median Size $m^2$ & Count \\
\midrule
Cork & 208 & 462 & 208 & 134 & 145 & 119 & 23 & \euro 3,330 & 95 & 1,299 \\
Dublin & 992 & 3,555 & 1,966 & 1,100 & 781 & 2,256 & 415 & \euro 4,960 & 95 & 11,065 \\
Galway & 82 & 193 & 49 & 39 & 39 & 122 & 37 & \euro 3,371 & 100 & 561 \\
Limerick & 41 & 236 & 52 & 43 & 38 & 96 & 24 & \euro 2,646 & 98 & 530 \\
Rural & 3,141 & 3,946 & 884 & 644 & 577 & 548 & 105 & \euro 2,214 & 109 & 9,845 \\
Towns & 1,142 & 2,772 & 602 & 503 & 347 & 633 & 159 & \euro 2,906 & 104 & 6,158 \\
\midrule
\textbf{Total} & \textbf{5,606} & \textbf{11,164} & \textbf{3,761} & \textbf{2,463} & \textbf{1,927} & \textbf{3,774} & \textbf{763} & \textbf{\euro 3,283} & \textbf{102} & \textbf{29,458} \\
Median Price per $m^2$ & \euro 2,647 & \euro 3,118 & \euro 3,800 & \euro 3,525 & \euro 3,267 & \euro 4,167 & \euro 3,158 &  &  &  \\
\bottomrule
\end{tabular}
\end{table}
\end{landscape}

\section{Methodology}\label{sec:Methodology}

\subsection{Geospatial Approach with GAM}\label{sec:GAM_definition}

We define the S-GAM, accounting for localised submarkets, and describe the knot selection process. For comparison, the N-GAM is defined in Section~\ref{NGAM}, which does not account for localised submarkets. Following the approach of \textcite{Hurley2022}, the natural log of price per $m^2$ is modelled as a random variable assuming Gaussian error in residuals. 

The S-GAM with localised submarket adjustment is specified as


\vspace{-1cm}
\begin{center}
\begin{flalign}\label{eqn:submarket_gam}
    \mathrm{log}(y_i) = \beta_0 + 
    \bm{Z}_{i}\bm{\beta} + \bm{P}_{i}\bm{\gamma}^{(\ell_i)} + 
    \sum_{j=1}^{4} f_j^{(\ell_i)}(x_{ij}; k_j) +
    f_5(s_{1i}, s_{2i}; k_5) +
    f_6(ec_i; k_6) + 
    \varepsilon_i,
     \hspace{1em}
\end{flalign}
\end{center}

where $y_i$ is the price per $m^2$ of observation $i$ with an iid residual value $\varepsilon_{i} \sim  N(0,\, \sigma^2)$. The intercept, $\beta_0$, corresponds to the expected value of $\text{log}(y_i)$, which is the average price per $m^2$ nationally. $\bm{Z}_i$ comprises the $p$ = 11 indicator descriptor variables outlined in Table \ref{tab:dummy_variables} for property $i$, \{$\mathds{1}_\textit{Attic Conversion}$, \dots, $\mathds{1}_\textit{New Property}$\}. $\bm{P}_i$ is comprised of $q$ = 7 dummy indicators for the property types outlined in Table \ref{tab:variables}, \{$\mathds{1}_\textit{Detached}$, \dots, $\mathds{1}_\textit{Duplex}$ \}. A sum to grand-mean constraint is imposed on $\bm{P}$ for identifiability. We allow for submarket-specific effects of property types through the associated regression coefficients varying by submarket, $\bm{\gamma}^ {(\ell_i)}$, where $\ell_i \in$ \{$\mathds{1}_\textit{Cork}$, \dots, $\mathds{1}_\textit{Towns}$\} is the submarket of observation $i$ (Table \ref{tab:area_desc}).

 The number of bedrooms, bathrooms, property size, and month of sale of observation $i$ are grouped as $\bm{X}_i = \{Beds_i, \: Baths_i, \: Size_i, \: Month_i\}$. The individual variables $x_{ij}$ are modelled using smoothing splines $f_j^{(\ell_i)}$ specific to the submarket of observation $i$, with a corresponding number of knots $k_j$.  Cubic regression splines are used for $Beds$, $Baths$ and $Size$, and a $p$ spline is used for $Month$, with 8, 7, 40 and 10 knots, respectively, the selection of which is described in Section \ref{subsubsec:gam_fitting}.

A GP smoothing spline $f_5$ is specified for the interaction between Cartesian coordinates, $(s_1, s_2)$, with 400 knots. The Eircode region of observation $i$, $ec_i$ is modelled using a MRF $f_6$ with 139 knots correspondingref to the number of Eircode regions, and a second-order neighbourhood structure. Both spatial components $f_5$ and $f_6$ are zero-centered. 

\subsubsection{Model Fitting and Knot Selection}\label{subsubsec:gam_fitting}

The model is fitted using the \textit{mgcv} package developed by \textcite{wood2017}, and parameters are estimated using penalised maximum likelihood. The number of knots for the smoothing splines for bedrooms and bathrooms are chosen as the maximum number of bedrooms and bathrooms observed in the dataset. 

\begin{center}
\begin{figure}[h!]
    \centering
    \includegraphics[width = 1\linewidth]{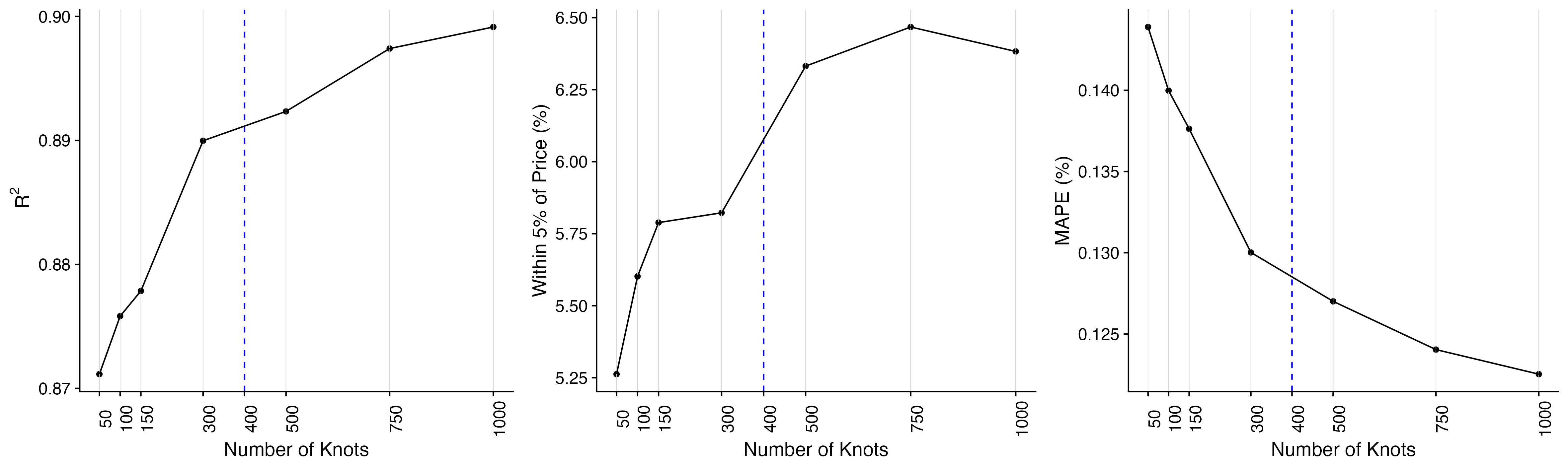}
    \caption{\small  Knot selection for Location showing the $R^2$ value, proportion of predictions within 5\% of the true values and the median absolute percentage error (MAPE)}
    \label{fig:location_knot_plots}
\end{figure}
\end{center}

Cross-validation is used to select the number of knots for the cubic regression splines corresponding to the \textit{size} variable and the GP location element. Figure \ref{fig:location_knot_plots} highlights the results from a 5-fold cross-validation for the GP spline using 8 potential knot values. The plots show the $R^2$ value, the proportion of predicted values within 5\% of the actual property price, and the median absolute percentage error (MAPE). \textcite{wood2017} recommends choosing the number of knots \textit{k} considering both computational efficiency and model accuracy. The optimal knot choice of 400 is highlighted with a blue dashed line, selected as an ``elbow" considering each metric, where performance is balanced with parsimony and computational effort. A similar process was used to select 40 knots for property \textit{size}.

\vspace{-1cm}

\newpage

\subsubsection{National GAM}\label{NGAM}

The N-GAM follows from the S-GAM in Equation \ref{eqn:submarket_gam} without the factor-by-spline interaction terms between property \textit{area} and \textit{size}, \textit{beds}, \textit{baths} and \textit{month}. Thus, there is one spline for each variable regardless of \textit{area}. Similarly, there is no interaction term between \textit{area} and \textit{property type}. The model simplifies to

\begin{center}
\begin{flalign}\label{eqn:national_gam}
      \mathrm{log}(y_i) = \beta_0 + 
    \bm{Z}_{i}\bm{\beta} + \bm{P}_{i}\bm{\gamma} + 
    \sum_{j=1}^{4} f_j(x_{ij}; k_j) +
    f_5(s_{1i}, s_{2i}; k_5) +
    f_6(ec_i; k_6) + 
    \varepsilon_i,
     \hspace{1em}
\end{flalign}
\end{center}
where all terms were defined previously.

\subsection{Hedonic Regression Model}

The hedonic regression model is used as a baseline comparison to the GAM approach. This multiple linear regression model assumes a linear relationship between property attributes and price and effectively fits a single equation to all observations. The variables in the multiple linear regression model are adequate for assessing the impacts of externalities on house prices \parencite{Oust2020}. Irish property studies have modelled dwelling prices based on the structural type; \textcite{Hurley2022} and \textcite{o2011constructing} specify property types as detached, semi-detached and bungalow, for example. Similarly, common internal property characteristics include the number of bedrooms and floor area \parencite{shinnick1997measuring}. More comprehensive property records support the inclusion of additional variables such as house age, living area, and plot size \parencite{Farber2006}.
\textcite{pace1998spatial} incorporate dummy variables for the presence of the property features, such as ``carport'', ``aluminium doors'', and ``aircon'', and \textcite{Hurley2022} use similar variables for ``attic conversion'', ``cul-de-sac'' and ``fireplace''. Distance variables predominantly consist of distance to the Central Business District (CBD) \parencite{Dubin1988, Hurley2022}, or distance to shopping facilities \parencite{Farber2006}. 

The parameters of the hedonic model are frequently estimated using ordinary least squares whereby the variables are assumed invariant across space and time \parencite{Farber2006}. Hence, this is a global regression model, which fits a single equation to the relationship between the independent variables and the dependent variable. This leads to difficulty in capturing spatial autocorrelation and spatial heteroscedasticity in the data, and the spatial dependence is captured either in the regressors or in the unstructured error term.
\textcite{Oust2020} conclude that the accuracy of the traditional hedonic model is vulnerable to the identification of housing attributes in addition to the effects of time and location. In a similar study, \textcite{Haan2013} identify a risk of omitted variable bias in the model when one or more relevant variables are excluded. \textcite{Basu1998} outline that similar values for omitted variables amongst neighbours lead to spatially correlated error terms.

Our hedonic regression model takes the form

\begin{center}
\begin{flalign}\label{eqn:hedonic}
    \mathrm{log}(y_i) = \beta_0 + 
    \bm{Z}_{i} \bm{\beta} + \bm{P}_i\bm{\gamma}+ 
 \bm{X}_{i} \bm{\kappa} + 
    \varepsilon_i,
\end{flalign}
\end{center}

where $\bm{Z}_i$, $\bm{P}_i$ and $\bm{X}_i$, with corresponding coefficients $\bm{\beta}$, $\bm{\gamma}$ and $\bm{\kappa}$, follow from Equation \ref{eqn:submarket_gam}. There is no interaction term between \textit{property type} and \textit{area}. Since this model is a comparison to geospatial modelling approaches, locational elements are not incorporated; thus, property coordinates, \textit{area}, and Eircode regions are not included in Equation \ref{eqn:hedonic}. The model coefficients are estimated using ordinary least squares.

\subsection{Random Forest Model}

RF is an ensemble learning technique comprised of multiple decision trees \parencite{breiman2001random}. Using a training set, each decision tree splits the feature space based on rules, or control sequences, at each node \parencite{hastie2009elements}. The leaf node at the end of the decision tree is reached when a stopping criterion is met, and a prediction is provided. The RF averages the predictions of many decision trees to prevent overfitting and improve model generalisation. Bootstrap sampling is used to randomly select samples and features from the training data at each split, further increasing model robustness. 

\begin{figure}[!t]
\centering
\setlength{\abovecaptionskip}{4pt}
\setlength{\belowcaptionskip}{0pt}
\includegraphics[width=0.82\linewidth]{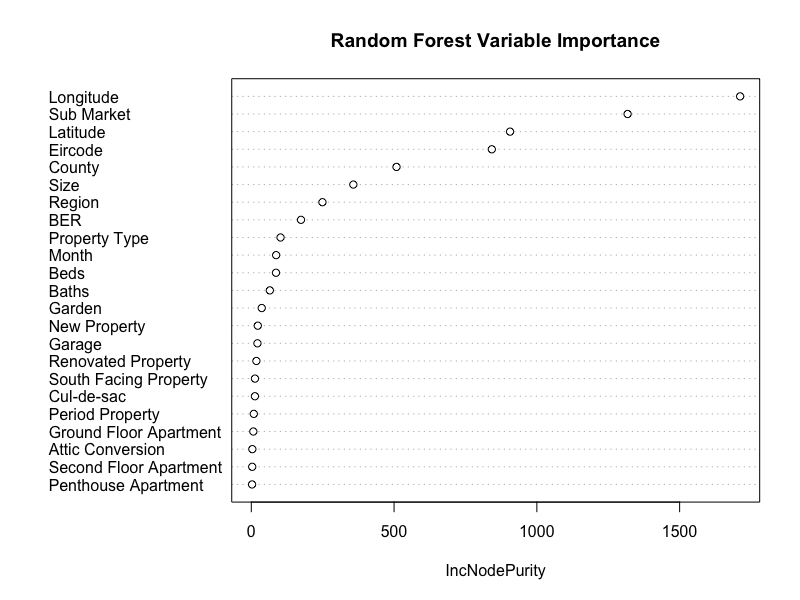}
\caption{\small Variable importance (increase in node purity) for the RF model}
\label{fig:rf_varImp}
\vspace{-0.4cm}
\end{figure}

The variables used to train our RF are the same as those of the S-GAM in Section \ref{sec:GAM_definition}. Additional address information, namely the county of each property and the region within the county, were included. The reason for including supplementary location variables relates to the ability of the RF to inherently manage variable importance and thus, variable selection. The final model was fitted using the default RF regression parameters, namely 500 trees with 7 variables randomly selected as candidates at each split. The model, fitted using the \textit{RandomForest} package by \textcite{pkg_randomForest}, provides insights into variable importance by assessing how much a feature improves tree splits across the entire forest. The locational factors, such as longitude, latitude, submarket and Eircode region, have the greatest variable importance in our RF (Figure \ref{fig:rf_varImp}). The sparse data in rural areas, as seen in Figure \ref{fig:subarea_map_lbls}, can prevent the RF from generalising predictions in such areas, resulting in unstable predictions and high variance. Despite its strengths as a predictive model, the RF lacks the interpretability of individual predictions and is often termed a ``black box" model \parencite{rudin2019stop}.

\subsection{Analysis}

The models described in Section \ref{sec:Methodology} are tested using 5-fold cross-validation. The training and test sets contain approximately 23,566 observations and 5,892 observations, respectively. The metrics used for model comparison are the $R^2$ values, MAPE, and root mean squared error (RMSE) values. The proportion of observations within their respective prediction intervals and the proportion of predicted values within specified intervals of each observation are also listed. The variable selection approach used is that of \textcite{gelman2007data}, where variables with expected signs are retained in the model. 

Moran's $I$ is used to assess the degree of spatial autocorrelation of model residuals—this is the Pearson's product-moment correlation coefficient, which measures the correlation of an observation with itself through the weights of a distance-based weight matrix \parencite[pp.~262-265]{getis2009spatial}. The values of $I$ lie within [-1,1], where a positive value signifies similarity amongst neighbours, a negative value indicates dispersion and a value close to 0 is associated with spatially independent observations \parencite{gaetan2010spatial}.

The performance of each model is analysed across submarkets, where the 5-fold cross-validation results are aggregated by submarket. Furthermore, we compare the predicted and actual prices for each model at both the national and submarket levels. 

The S-GAM is interpreted in detail. First, the coefficients are plotted and described, including relative scalings for the sum-to-zero encoded factor variables, smoothing splines, and a fusion of spatial components. The corresponding parametric coefficient estimates with respective 95\% confidence intervals are also reported.

\section{Results}\label{sec:results}

\subsection{Fitted Models Results}

In this section, we present the results of the N-GAM, S-GAM, hedonic regression and random forest model. The results from the 5-fold cross-validation are averaged for each model and presented in Table \ref{tab:cv_all}. 

\begin{table}[t!]
\centering
\caption{\label{tab:cv_all}Results from 5-fold cross validation}
\small 
\begin{tabular}{@{}%
    >{\raggedright\arraybackslash}p{0.15\linewidth}%
    >{\raggedleft\arraybackslash}p{0.05\linewidth}%
    >{\raggedleft\arraybackslash}p{0.08\linewidth}%
    >{\raggedleft\arraybackslash}p{0.08\linewidth}%
    >{\raggedleft\arraybackslash}p{0.08\linewidth}%
    >{\raggedleft\arraybackslash}p{0.08\linewidth}%
    >{\raggedleft\arraybackslash}p{0.08\linewidth}%
    >{\raggedleft\arraybackslash}p{0.08\linewidth}%
    >{\raggedleft\arraybackslash}p{0.08\linewidth}@{}}
\toprule
 & $R^2$ & RMSE & MAPE & Within 5\% of Price & Within 10\% of Price & Within 50\% PI & Within 95\% PI & Moran's I\\
\midrule
Hedonic Model & 0.66 & \euro 155,573 & 0.25 & 2.8\% & 27.9\% & 53.1\% & 95.7\% & 0.21\\
\addlinespace
N-GAM & 0.84 & \euro 104,716 & 0.15 & 5.8\% & 50.7\% & 46.1\% & 92.5\% & 0.05\\
\addlinespace
S-GAM & 0.85 & \euro 101,622 & 0.14 & 6.2\% & 55.5\% & 46.6\% & 92.9\% & 0.03\\
\addlinespace
Random Forest & 0.87 & \euro 97,119 & 0.14 & 7.1\% & 55.5\% & 58.6\% & 97.3\% & 0.01\\
\bottomrule
\multicolumn{9}{@{}p{\linewidth}@{}}{\textit{Note:} RMSE is the root mean squared error; MAPE is the median absolute percentage error; PI is the prediction interval.}\\
\end{tabular}
\end{table}

The hedonic model, which has no locational attributes, represents a benchmark for assessing the spatial enhancements of the two GAMs and the RF. The N-GAM, which incorporates non-linear relationships and spatial flexibility, outperforms the hedonic model with an \rsq value of 0.84 over 0.66. Similarly, the MAPE is reduced to 0.15 for the N-GAM, compared to 0.25 for the hedonic model. 

The S-GAM, incorporating submarket segmentation into the N-GAM, performs relatively similarly to the N-GAM, with an \rsq value of 0.85 and reduced MAPE of 14\%. The minute differences in \(R^2\), RMSE, and MAPE between the N-GAM and S-GAM are likely due to the similarities in model structure, where the inclusion of submarkets is most intended for the interpretability of results. \textcite{Hurley2022} report similar results when comparing a hedonic model to a flexible geospatial GAM in Dublin, with $R^2$ values of 0.73 and 0.87, respectively. 

The RF has the greatest \rsq value (0.87) and the lowest RMSE value of all models. This is an expected result of such a machine learning approach trained on a vast amount of data. The RF has a similar MAPE to both of the GAM approaches, suggesting all models have similar average error values. Similar results were found by \textcite{Hurley2022}, who trained a RF with an $R^2$ value of 0.87 and a MAPE value of 10.92\%. The increased MAPE value in our case is likely due to the heterogeneity of observations at a national scale. The RF and S-GAM have an equal proportion (55.5\%) of predictions within 10\% of their true price, which is greater than that of the N-GAM (50.7\%), and hedonic model (27.9\%). Considering the proportion of values within 5\% of their true values, the RF (7.1\%) outperforms the S-GAM (6.2\%). 

The 50\% and 95\% prediction intervals of the GAM models are both narrower than expected; the S-GAM has 46.6\% and 92.9\% of predictions within their respective 50\% and 95\% prediction intervals. In comparison, the prediction intervals of the RF, which are calculated using a bootstrapping approach, are overly conservative. In this case, 58.6\% and 97.3\% of the true values are contained within their 50\% and 95\% prediction intervals, respectively. The wider prediction intervals are attributed to the RF's inability to achieve good uncertainty estimates \parencite{breiman2001random}. 

The residuals of the hedonic model demonstrate a high level of positive spatial autocorrelation, with a Moran's $I$ value of 0.21. Similarly, \textcite{Oust2020} report a $I$ value of 0.14 for an ordinary regression model with submarkets. Increasing the spatial complexity by including the coordinates, Eircode, and submarket of each property in the S-GAM effectively accounts for the spatial variation between observations, reducing the $I$ value to 0.03.

\begin{center}
\begin{figure}[b!]
    \centering
    \includegraphics[width = 1\linewidth]{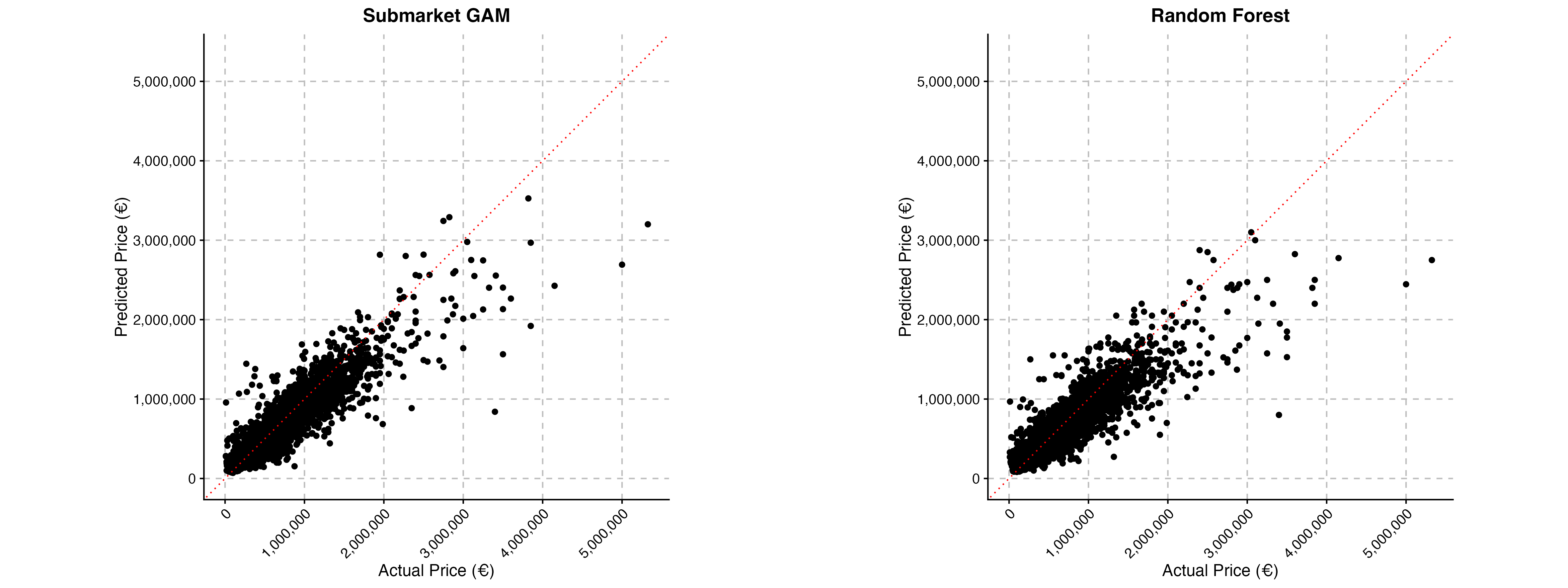}
    \caption{\small Predicted vs. actual price of the S-GAM and RF with perfect prediction represented in red}
    \label{fig:pred_act_plots}
\end{figure}
\end{center}

The predicted and actual values are plotted for the S-GAM and RF in Figure \ref{fig:pred_act_plots}. Both models have a similar behaviour, with points following the red line representing perfect prediction, and increased variance for larger values. Both models appear to underestimate properties with actual price values over \euro3,000,000. This is likely due to the low quantity of such properties in the training data and the lack of property attributes to describe higher-valued properties. 

The performance of each model across submarkets is presented in Table \ref{tab:cv_submarkets}. The RF has reduced accuracy in rural areas with an \rsq value of 0.52, compared to the S-GAM with an \rsq value of 0.70. The difference in model performance is also apparent in the RMSE and MAPE values which are greater for the RF. The S-GAM has increased precision in rural areas, with 48.6\% and 5.6\% of predictions within 50\% and 5\% of the true price, compared to the RF, which has 13.1\% and 1.3\% of predictions within 50\% and 5\% of their true price, respectively. 

\begin{figure}[!t]
\centering
\setlength{\abovecaptionskip}{4pt}
\setlength{\belowcaptionskip}{0pt}
\includegraphics[width=0.82\linewidth]{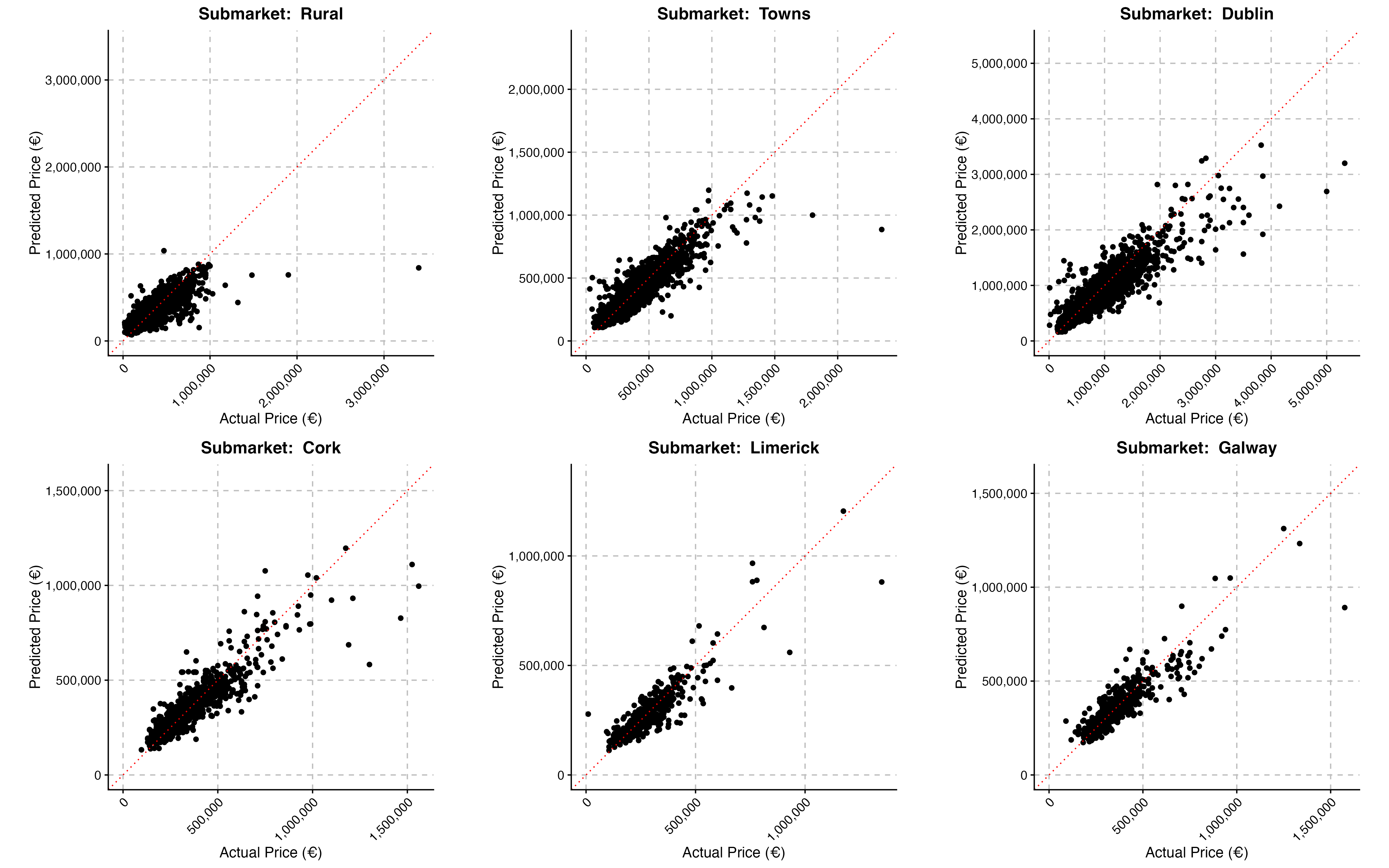}
\caption{\small Predicted vs. actual price of the S-GAM across submarkets, with perfect prediction represented in red}
\label{fig:g2_subareas_pred_act}
\vspace{-0.4cm}
\end{figure}

\begin{figure}[!t]
\centering
\setlength{\abovecaptionskip}{4pt}
\setlength{\belowcaptionskip}{0pt}
\includegraphics[width=0.82\linewidth]{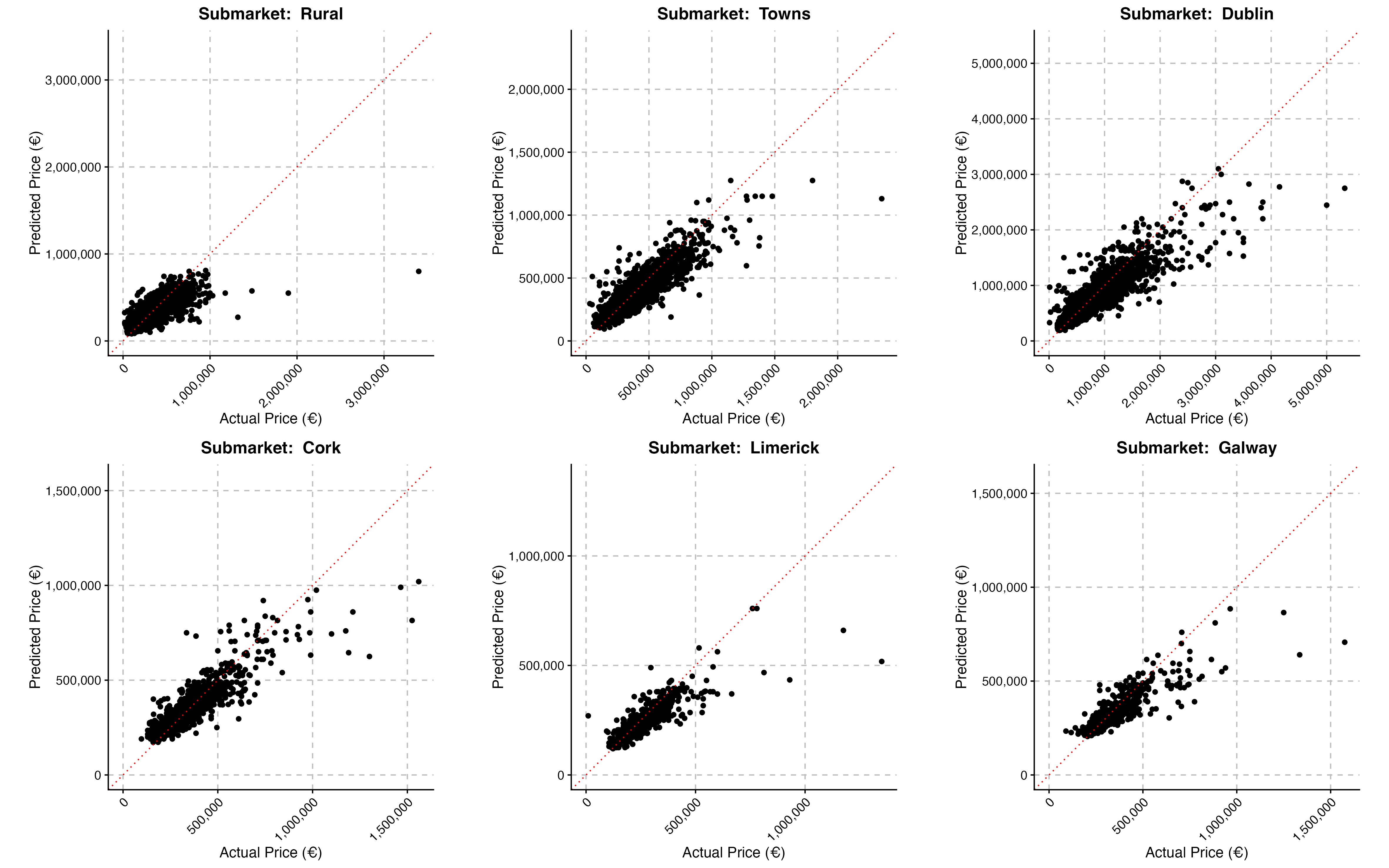}
\caption{\small Predicted vs. actual price of the RF across submarkets, with perfect prediction represented in red}
\label{fig:rf_subareas_pred_act}
\vspace{-0.4cm}
\end{figure}

The predicted values are plotted across submarkets for the S-GAM in Figure \ref{fig:g2_subareas_pred_act}, and for the RF in Figure \ref{fig:rf_subareas_pred_act}. Both models demonstrate heteroscedasticity in each submarket, where there is increased variance for larger values of property prices. 
In rural areas, the S-GAM highly underestimates a single property with a sale price of \euro3,000,000. This property listing contains no text description and thus could not be text-mined for characteristics or erroneous measures of \textit{size}.

\vspace{-0.8cm}

The S-GAM outperforms the RF in Galway, Limerick, Dublin, Towns and Rural areas; both models have similar performance metrics in Cork. In Figure \ref{fig:g2_subareas_pred_act}, the S-GAM appears to overestimate the price of a property in Limerick; this is likely a reason for the MAPE of 29\% in Table \ref{tab:cv_submarkets}, which is greater than the MAPE values of the S-GAM in other submarkets. The N-GAM and S-GAM perform similarly across all submarkets. There is a notable difference in the \rsq values of each model for rural areas. Since the RMSE, MAPE, precision, and predictive interval metrics remain similar, this is likely an artefact of the cross-validation approach. 

The RF model and S-GAM have improved performance in Dublin compared to other submarkets, with \rsq values of 0.82 and 0.84, respectively. This could be attributed to the increased number of observations in Dublin, as observed in Figure \ref{fig:subarea_map_lbls}. The S-GAM is selected as the best model due to the increased performance within submarkets compared to the RF and N-GAM. While the RF has improved performance at a national scale, this approach lacks interpretability and reliable uncertainty estimates \parencite{breiman2001random}.

\begin{landscape}\begin{table}
\centering
\caption{\label{tab:cv_submarkets}Results from 5-fold cross-validation within submarkets}
\centering
\begin{tabular}[t]{%
    p{0.25\linewidth} 
    p{0.08\linewidth} 
    p{0.05\linewidth} 
    >{\raggedright\arraybackslash}p{0.05\linewidth} 
    >{\raggedleft\arraybackslash}p{0.08\linewidth} 
    >{\raggedright\arraybackslash}p{0.08\linewidth} 
    >{\raggedright\arraybackslash}p{0.08\linewidth} 
    >{\raggedright\arraybackslash}p{0.08\linewidth} 
    >{\raggedright\arraybackslash}p{0.08\linewidth}} 
\toprule
Model & Submarket & $R^2$ & RMSE & MAPE & Within 5\% of Price & Within 10\% of Price & Within 50\% Prediction Interval & Within 95\% Prediction Interval\\
\midrule
 & Rural & 0.47 & \euro 93,457 & 0.26 & 2.9\% & 28.1\% & 52.3\% & 94.3\%\\

 & Towns & 0.48 & \euro 110,821 & 0.25 & 2.8\% & 27\% & 51.5\% & 95.6\%\\

 & Dublin & 0.60 & \euro 218,244 & 0.25 & 2.4\% & 25.1\% & 50.6\% & 96.6\%\\

 & Cork & 0.64 & \euro 91,398 & 0.16 & 4.7\% & 43.6\% & 75.1\% & 97.8\%\\

 & Limerick & 0.75 & \euro 64,364 & 0.21 & 5.8\% & 45.3\% & 75.1\% & 99\%\\

\multirow[t]{-6}{*}{\raggedright\arraybackslash Hedonic Regression} & Galway & 0.61 & \euro 96,049 & 0.18 & 4.3\% & 34.9\% & 65.6\% & 98.9\%\\
\cmidrule{1-9}
 & Rural & 0.78 & \euro 59,166 & 0.16 & 5.4\% & 46.7\% & 44.4\% & 92\%\\

 & Towns & 0.81 & \euro 61,741 & 0.14 & 5.1\% & 55\% & 48\% & 93.9\%\\

 & Dublin & 0.82 & \euro 148,442 & 0.14 & 5.2\% & 52\% & 44.9\% & 92.3\%\\

 & Cork & 0.70 & \euro 80,188 & 0.13 & 6\% & 57.2\% & 48.7\% & 89.6\%\\

 & Limerick & 0.72 & \euro 78,857 & 0.39 & 5.4\% & 47.4\% & 42.2\% & 88.5\%\\

\multirow[t]{-6}{*}{\raggedright\arraybackslash N-GAM} & Galway & 0.65 & \euro 122,484 & 0.18 & 3.1\% & 32.4\% & 47.5\% & 89.2\%\\
\cmidrule{1-9}
 & Rural & 0.70 & \euro 74,367 & 0.16 & 5.6\% & 48.6\% & 44.8\% & 91.8\%\\

 & Towns & 0.83 & \euro 60,879 & 0.13 & 6\% & 57\% & 47.2\% & 93.3\%\\

 & Dublin & 0.84 & \euro 137,004 & 0.13 & 5.8\% & 55\% & 46.4\% & 92.9\%\\

 & Cork & 0.73 & \euro 76,242 & 0.12 & 6.2\% & 57.1\% & 47.6\% & 91.2\%\\

 & Limerick & 0.73 & \euro 81,756 & 0.29 & 5.7\% & 51.2\% & 43.2\% & 89.9\%\\

\multirow[t]{-6}{*}{\raggedright\arraybackslash S-GAM} & Galway & 0.72 & \euro 96,060 & 0.15 & 4.8\% & 43.3\% & 47.8\% & 89.9\%\\
\cmidrule{1-9}
 & Rural & 0.52 & \euro 142,004 & 0.56 & 1.3\% & 13.1\% & 29.2\% & 94.7\%\\

 & Towns & 0.71 & \euro 108,935 & 0.33 & 2.8\% & 23.5\% & 46.6\% & 98.2\%\\

 & Dublin & 0.82 & \euro 151,883 & 0.14 & 5.8\% & 50.7\% & 70.1\% & 99.3\%\\

 & Cork & 0.74 & \euro 77,950 & 0.16 & 4.7\% & 46.2\% & 71.1\% & 99.6\%\\

 & Limerick & 0.70 & \euro 94,719 & 0.40 & 1.5\% & 15.8\% & 39.5\% & 97.9\%\\

\multirow[t]{-6}{*}{\raggedright\arraybackslash Random Forest} & Galway & 0.64 & \euro 90,882 & 0.17 & 3.3\% & 42.5\% & 72\% & 99.8\%\\
\bottomrule
\multicolumn{9}{l}{\rule{0pt}{1em}{Note: RMSE is the root mean squared error,\             MAPE is the median absolute percentage error.}}\\
\end{tabular}
\end{table}
\end{landscape}

\subsection{Interpretation of Coefficients}

\begin{figure}[b!]
\centering
    \centering
    \includegraphics[width = 1\linewidth]{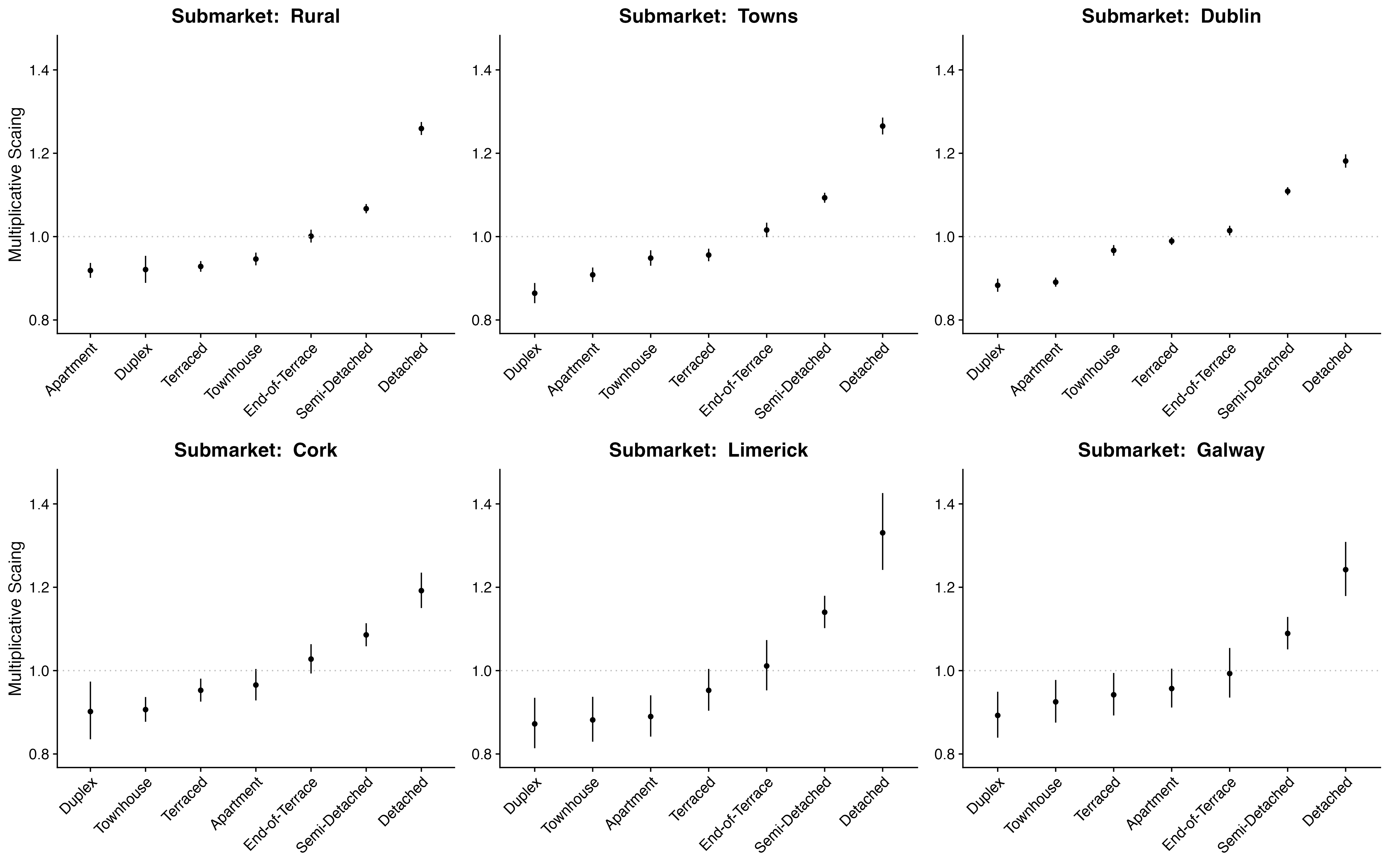}
    \caption{\small  Relative scalings of Property Types across submarkets}
    \label{fig:property_type_rel_scalings}
\end{figure}

In the following, we focus on the output of the S-GAM model. We note that when spatial random effects are included in the model, some care is warranted in interpreting regression coefficients, particularly when covariates exhibit spatial structure \parencite{Zimmerman03042022}. The levels of \textit{property type} are analysed across submarkets in Figure \ref{fig:property_type_rel_scalings}, sum-to-zero encoding ensures each level can be compared to the grand mean, or mean of levels within each submarket. The levels of each \textit{property type}, along with their 95\% confidence intervals, are plotted as relative scalings from the grand mean for each submarket, which are represented as grey dotted lines. In each submarket, detached and semi-detached properties have the greatest increase on the grand mean, while duplex properties, apartments and townhouses have the greatest decrease on the grand mean. The multiplicative scalings of \textit{property type} demonstrate a similar structure across submarkets, suggesting a uniform national structure. The relative scalings of \textit{property type} in Dublin align with those reported by \textcite{Hurley2022}. The premium of a detached property is lower in Dublin compared to rural areas and towns. This is likely due to the higher mean value of properties in Dublin (\euro 542,400)  compared to that in rural areas (\euro  269,100)  and towns (\euro  334,100).

\newpage

\begin{figure}[t!]
\centering
\setlength{\abovecaptionskip}{4pt} 
\setlength{\belowcaptionskip}{0pt}
\includegraphics[width=0.8\linewidth]{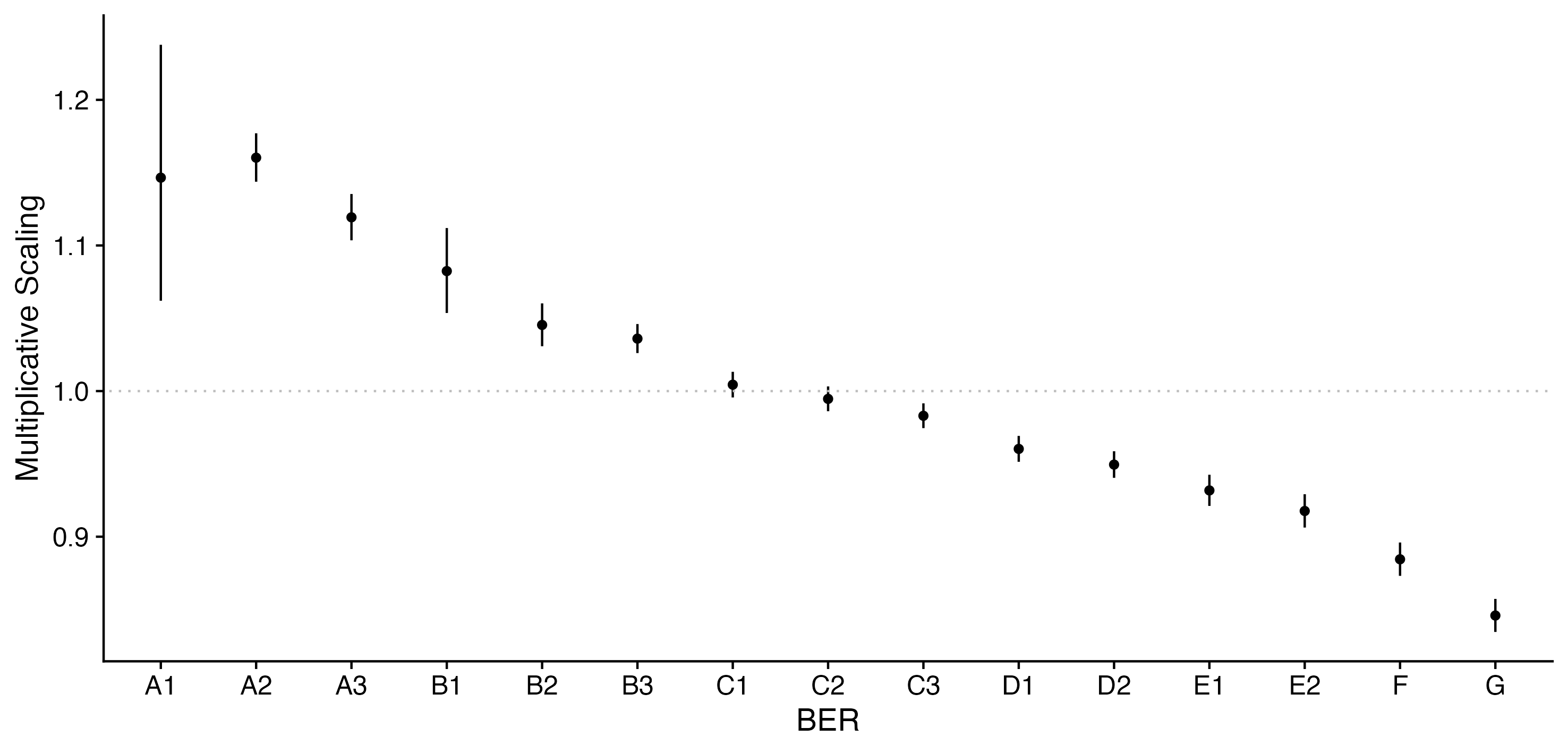}
\caption{\small Relative scalings of Building Energy Rating (BER) values}
\label{fig:ber_rel_scalings}
\vspace{-0.4cm} 
\end{figure}

The relative scalings of the levels of BER are plotted with their 95\% confidence intervals in Figure \ref{fig:ber_rel_scalings}, and a grey dotted line representing the grand mean is plotted again. The relative scalings of BER behave as expected, where more energy-efficient homes (such as A and B ratings) attain a higher premium over the grand mean, and less energy-efficient homes (such as E and F ratings) have a reduced scaling on the grand mean. Properties with a C rating appear to align with the grand mean across energy ratings; this is the most common energy rating of properties in Ireland \parencite{cso_ber2022}.

\subsection{Interpretation of Smooths}

\begin{figure}[h!]
\centering
\setlength{\abovecaptionskip}{4pt}
\setlength{\belowcaptionskip}{0pt}
\includegraphics[width=0.7\linewidth]{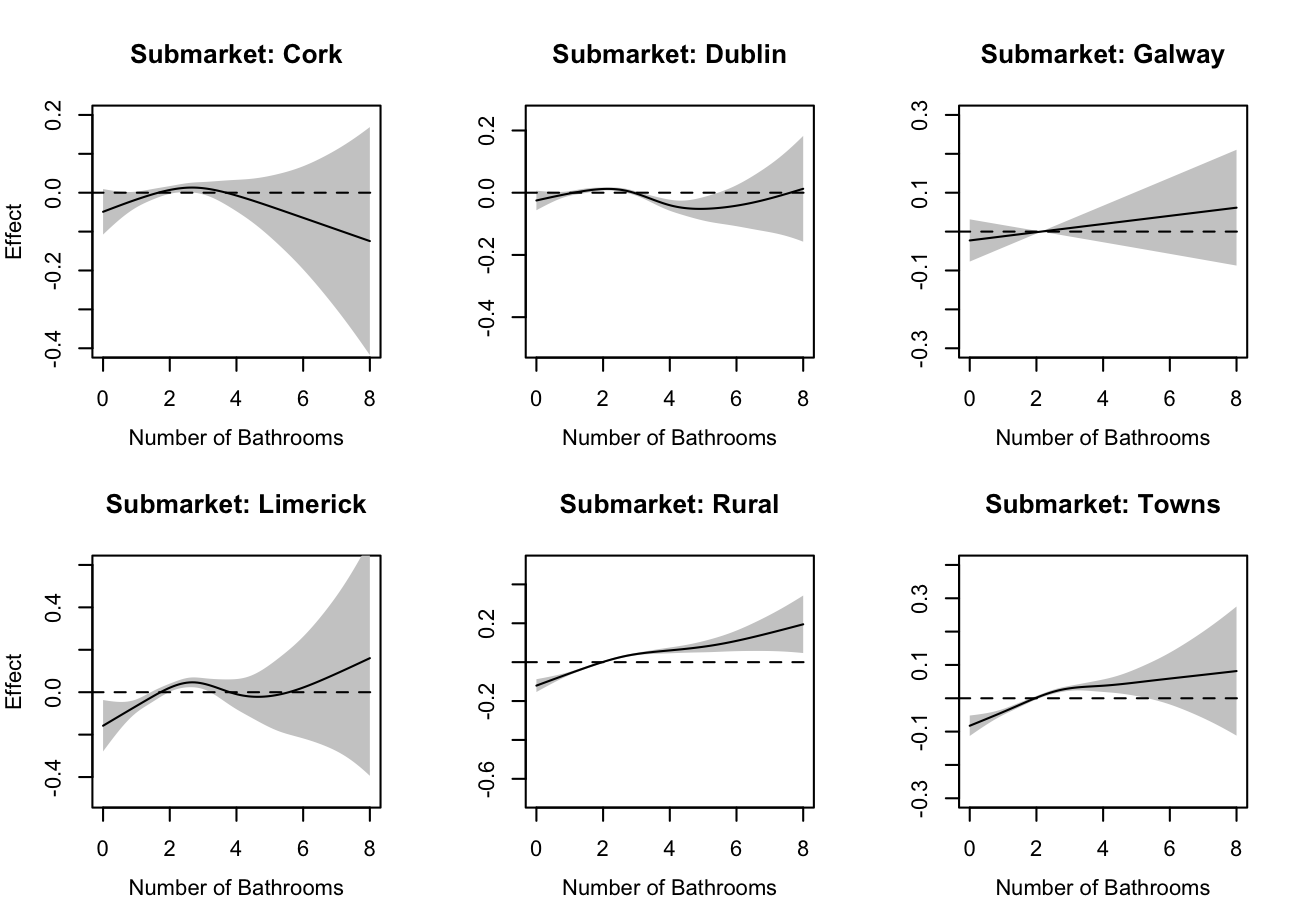}
\caption{\small Smoothing splines for the number of bathrooms across submarkets}
\label{fig:baths_splines}
\end{figure}

The cubic regression splines for the number of bathrooms within each submarket are plotted in Figure \ref{fig:baths_splines}, along with a shaded 95\% confidence interval. Each submarket demonstrates a positive effect of the addition of a first and second bathroom; the greatest effect is apparent in rural areas. A linear effect is apparent in Galway, while rural areas and towns demonstrate a similar levelling-off for bathrooms between 2 and 4, suggesting that the addition of a \nth{3} or \nth{4} bathroom has a diminishing effect. The reduced effect of a \nth{3} or \nth{4} bathroom in Limerick and Dublin is likely related to the characteristics of such properties. In Cork, this negative effect continues for all properties with over 3 bathrooms. Greater uncertainty is likely for over 4 bathrooms in Limerick and Cork due to the limited number of such properties in these submarkets. 

\begin{figure}[!t]
\centering
\setlength{\abovecaptionskip}{4pt}
\setlength{\belowcaptionskip}{0pt}
\includegraphics[width=0.82\linewidth]{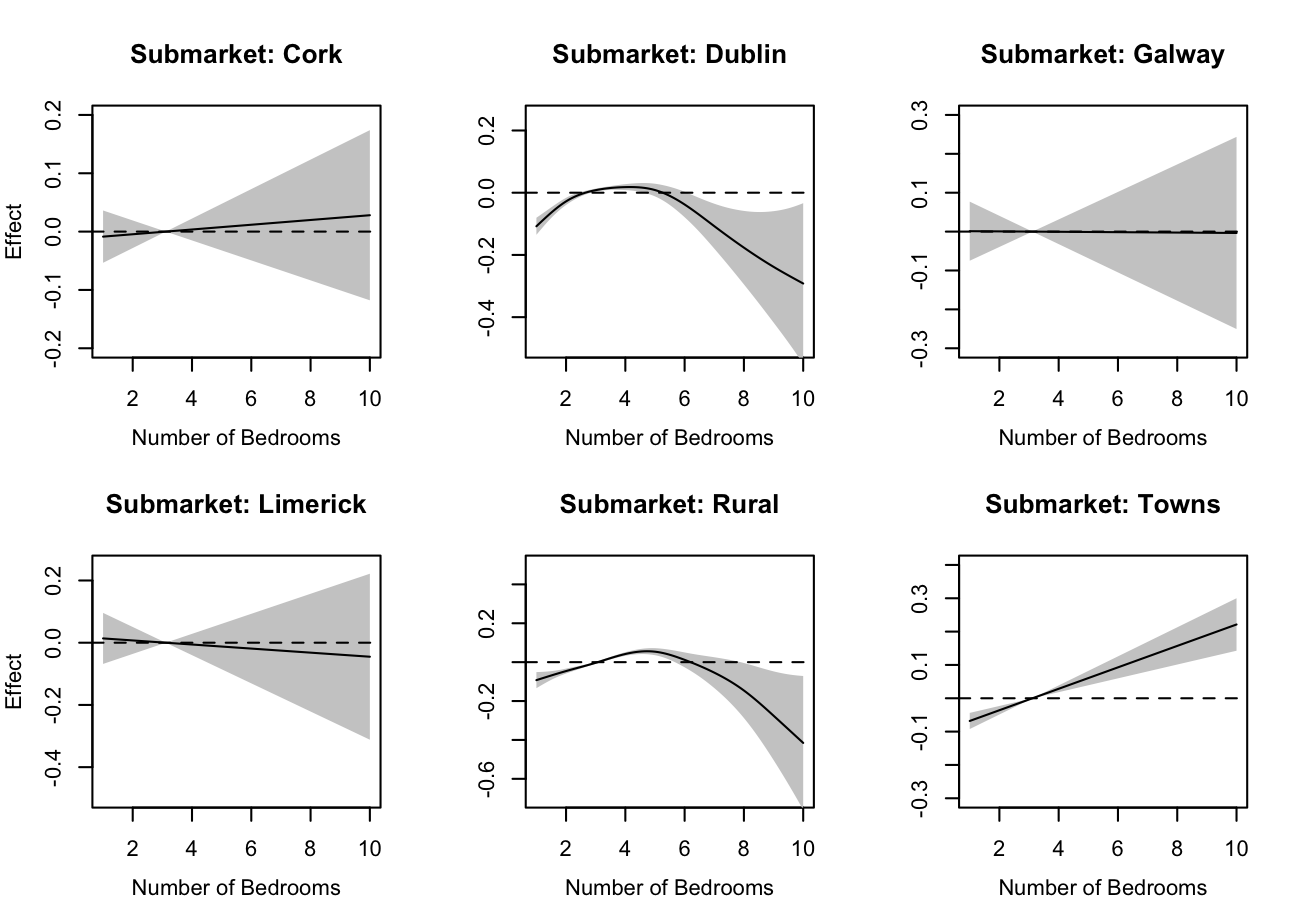}
\caption{\small Smoothing splines for the number of bedrooms across submarkets}
\label{fig:beds_splines}
\vspace{-0.4cm}
\end{figure}

Additional linear relationships are apparent in the smoothing splines for the number of bedrooms, shown in Figure \ref{fig:beds_splines}. The splines in Dublin and rural areas have a similar negative effect for larger properties; in contrast, the addition of a bedroom has a positive linear effect on property price in towns. 

\begin{figure}[b!]
\centering
\setlength{\abovecaptionskip}{4pt} 
\setlength{\belowcaptionskip}{0pt} 
\includegraphics[width=0.7\linewidth]{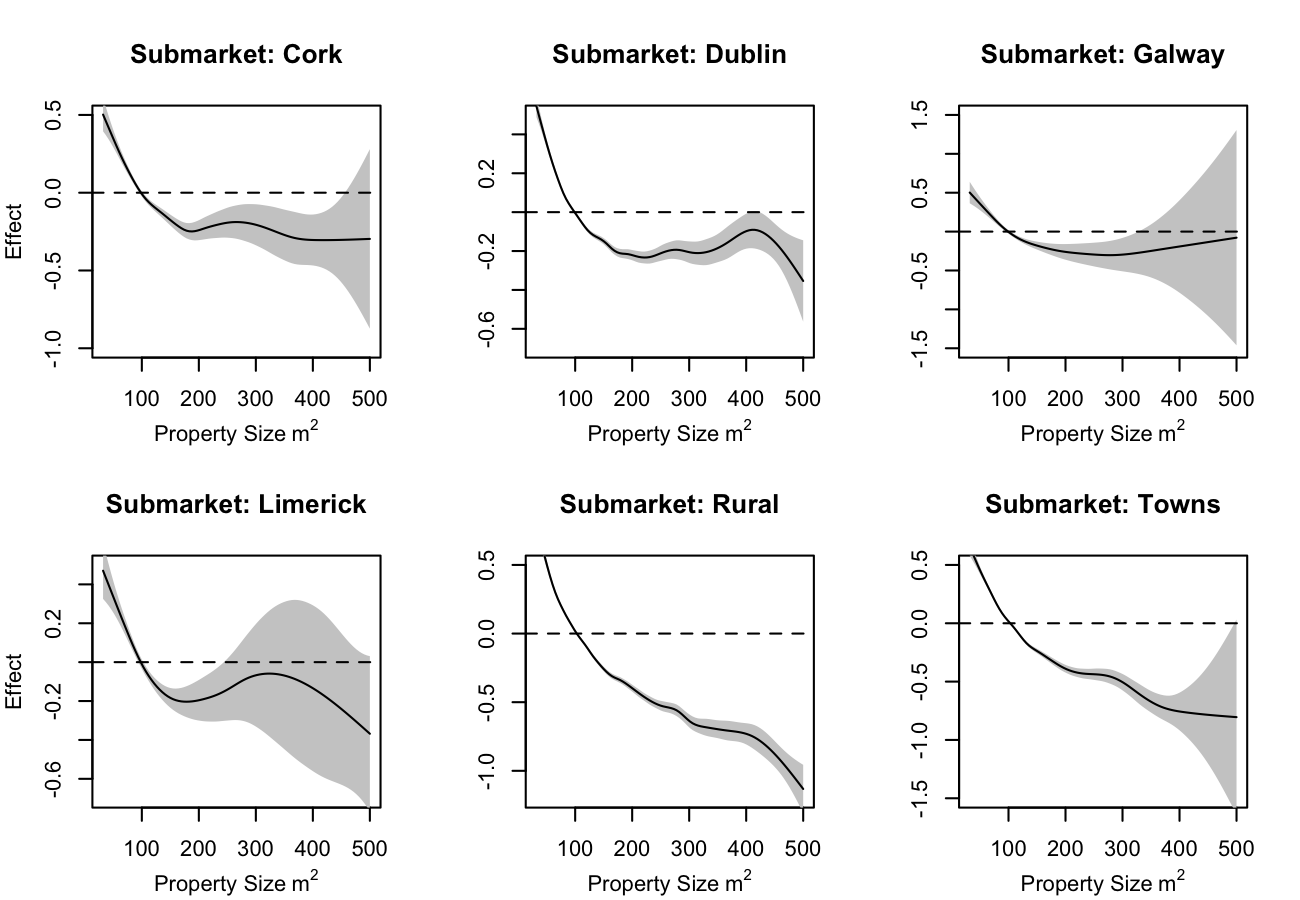}
\caption{\small Smoothing splines for property size across submarkets}
\label{fig:size_splines}
\vspace{-0.3cm} 
\end{figure}

The splines for property \textit{size} are plotted for each submarket in Figure \ref{fig:size_splines}. In rural areas and towns, increased property size appears to have a negative effect on property size; there is a plateau apparent for rural properties between 300$m^2$ and 400$m^2$, and town properties between 200$m^2$ and 300$m^2$. Dublin properties demonstrate a similar negative effect with a greater plateau for properties between 200$m^2$ and 350$m^2$, and an apparent positive effect for 400$m^2$ properties. \textcite{Hurley2022} identified a similar ``bump'' at 300$m^2$ caused by the prominence of large period properties in affluent areas. The effect of \textit{size} on price demonstrates differences in Cork, Galway and Limerick, where there is a slight positive effect at 200$m^2$ in Cork, a greater positive effect at 300$m^2$ in Limerick, and a more gradual increase from 300$m^2$ in Galway.

\begin{figure}[t!]
\centering
\setlength{\abovecaptionskip}{4pt} 
\setlength{\belowcaptionskip}{0pt} 
\includegraphics[width=0.7\linewidth]{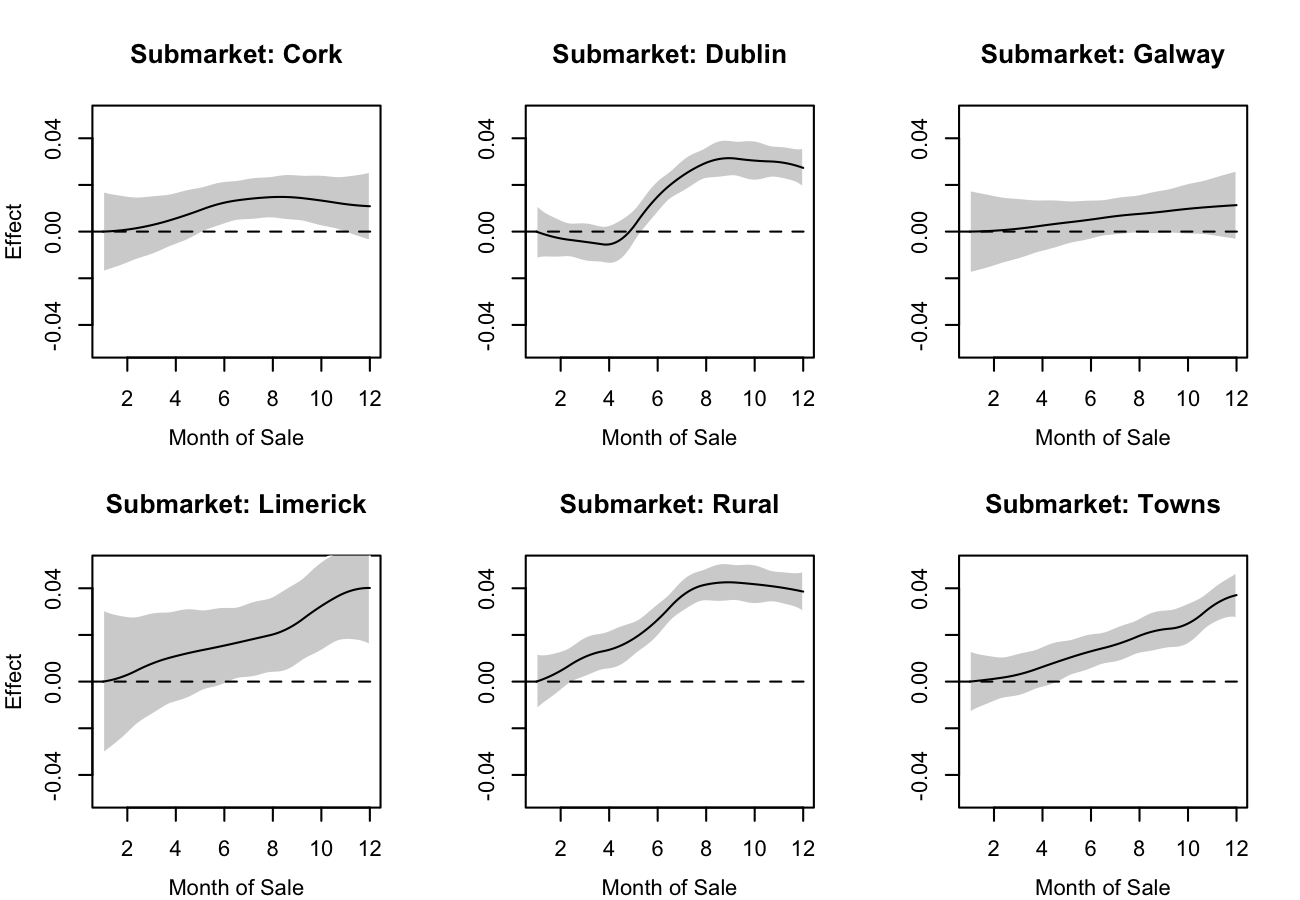}
\caption{\small Smoothing splines for month of sale across submarkets, centred at the origin for interpretability}
\label{fig:month_splines}
\vspace{-0.3cm} 
\end{figure}

\begin{figure}[!t]
\centering
\setlength{\abovecaptionskip}{4pt}
\setlength{\belowcaptionskip}{0pt}
\includegraphics[width=0.82\linewidth]{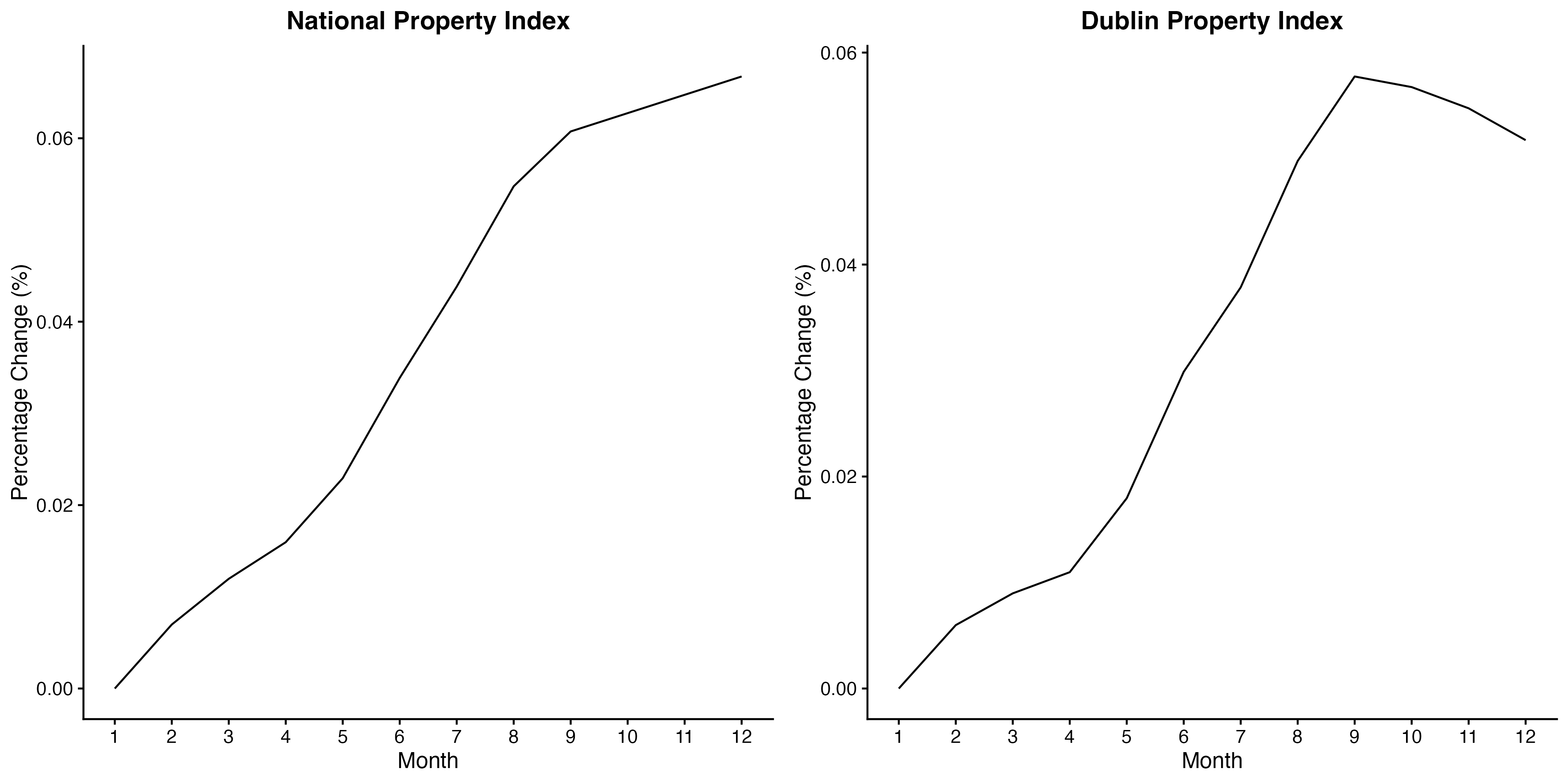}
\caption{\small Aggregated monthly inflation from the Residential Price Index}
\label{fig:cso_inflation_plots}
\vspace{-0.4cm}
\end{figure}

The temporal effects are captured by fitting $p$ splines to the month of sale within each submarket; these are plotted in Figure \ref{fig:month_splines}. Each submarket demonstrates an expected inflation in price throughout the year. In Limerick and Towns, there is a linear relationship between time and property value. The inflation is not as high in Cork and Galway, with a reduced linear relationship, and a slight plateau in the second half of the year in Cork. In Dublin, there is an initial negative temporal effect on property value for the first third of the year, followed by an increase to approximately 3\% during the second third of the year and a plateau for the remainder of the year. In rural areas, there is a similar increase to 4\% during the first two-thirds of the year, followed by a plateau. The Residential Property Index monthly inflation figures are used as a robustness check for the temporal effects of the S-GAM. These monthly figures, available for national and Dublin properties only, are aggregated to create month-on-month inflation plots in Figure \ref{fig:cso_inflation_plots}. The sharpest increase in inflation occurs in the \nth{4} month in both the monthly splines and the national reported figures. The monthly splines for Dublin and Rural submarkets most closely resemble the true inflation in Figure \ref{fig:cso_inflation_plots}, likely due to the greater number of observations in these areas compared to other submarkets. 

\begin{figure}[!t]
\centering
\setlength{\abovecaptionskip}{4pt} 
\setlength{\belowcaptionskip}{0pt} 
\setlength{\intextsep}{6pt}        

\begin{minipage}[t]{0.48\textwidth}
    \centering
    \includegraphics[width=0.8\linewidth]{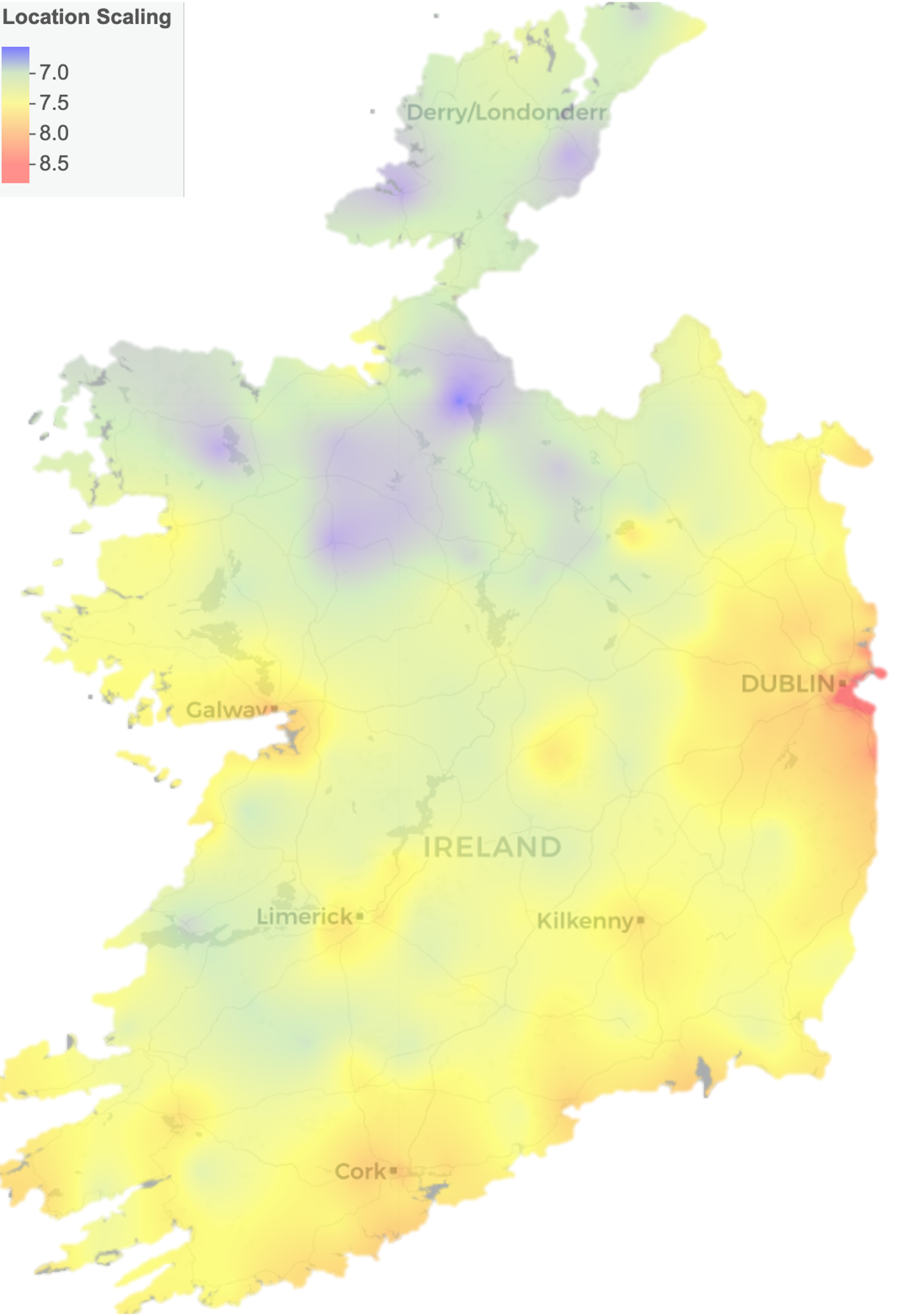}
    \caption{\small Map of the Gaussian process surface from the S-GAM}
    \label{fig:g2_GP_map_clipped}
\end{minipage}
\hfill
\begin{minipage}[t]{0.48\textwidth}
    \centering
    \includegraphics[trim={4cm 2cm 0 0},clip,width=1.2\linewidth]{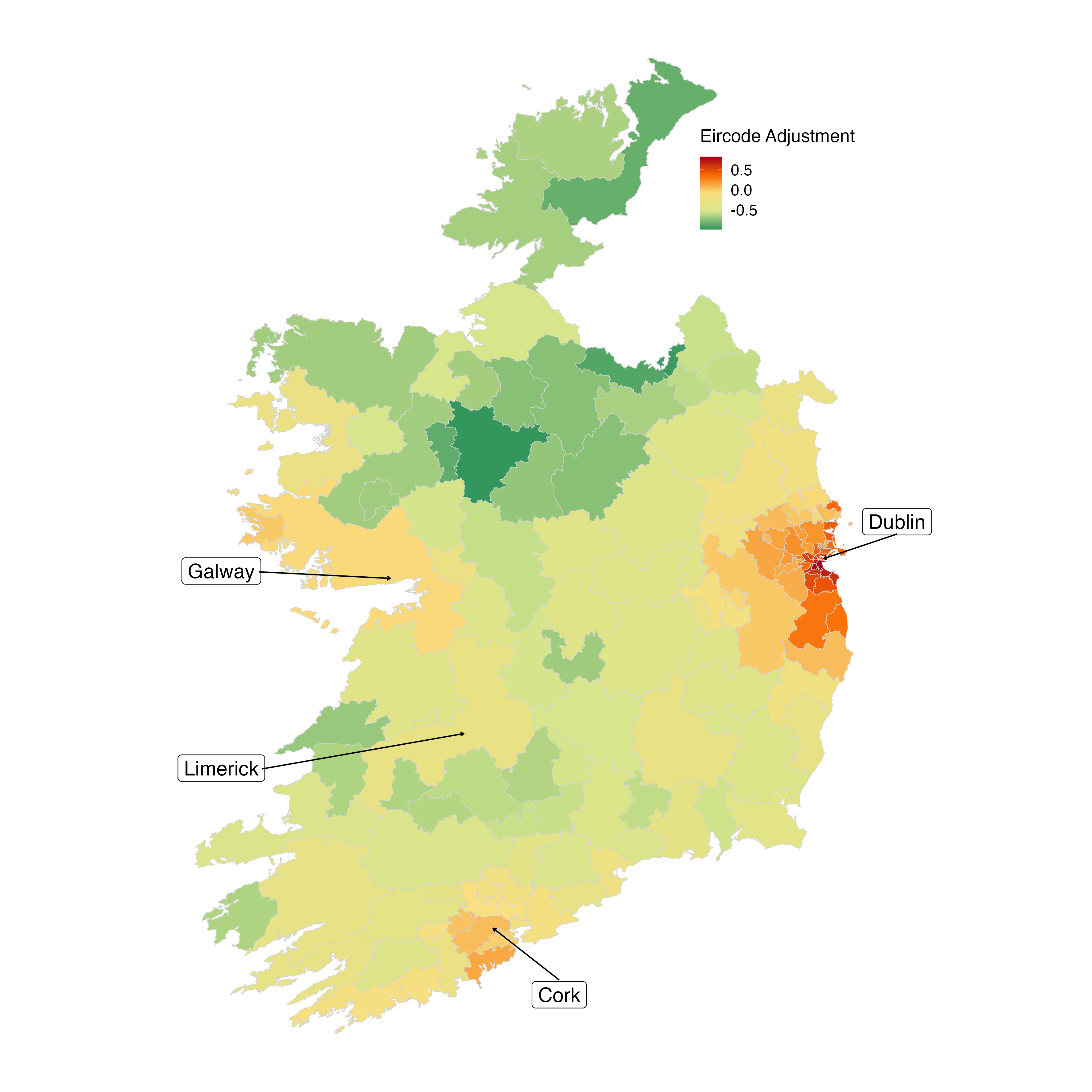}
    \caption{\small Map of the MRF Eircode values from the S-GAM}
    \label{fig:g2_MRF_map_lbls}
\end{minipage}

\end{figure}

The coefficients of dummy variables representing property features are outlined in the top rows of Table \ref{tab:tab:g2_coefficients}. The penthouse apartment has the greatest multiplicative increase of 1.15 in property price. Having a period property has a similar increase of 1.11 in property value. These large values are likely due to the affluent nature of such properties. The presence of a garage or a garden both increases property value; these attributes may be acting as a proxy for space or site size in the model. Similarly, the price increase associated with renovated and new properties likely serves as a proxy for property condition.

\subsection{Investigating Spatial Components}

The GP spatial surface is extracted and plotted on a log scale in Figure \ref{fig:g2_GP_map_clipped}. Dublin demonstrates the greatest effect on property prices, and this effect spreads to the surrounding areas. The cities of Cork, Galway and Limerick are captured as hot spots, suggesting positive effects of living in such areas. Living in proximity to the coast in the west, south or east of Ireland appears to increase property value. The lowest effects are in the midlands, areas east of Limerick, and the north—this aligns with the median Eircode region values plotted in Figure \ref{fig:map_median_ecode}. The areas with the lowest location scaling, represented as a dark blue, are spots in the northwest. These are areas with a low density of observations, as seen previously in Figure \ref{fig:subarea_map_lbls}. 

The MRF Eircode region adjustment values are plotted in Figure \ref{fig:g2_MRF_map_lbls}, with a log scale and labels for the city locations. Negative adjustments are apparent in the midlands and north, and positive adjustments are apparent in the south and east. The MRF estimation method results in neighbouring Eircode regions demonstrating similar values, thus creating a smooth map. The GP surface in Figure \ref{fig:g2_GP_map_clipped} captures the majority of the spatial effects, while the Eircode adjustments capture the macro-scale variation in property price due to Eircode values, and potentially demonstrate a bias towards certain Eircodes. 

\begin{center}
\begin{figure}[!h]
    \centering
    \includegraphics[width = 0.65\linewidth]{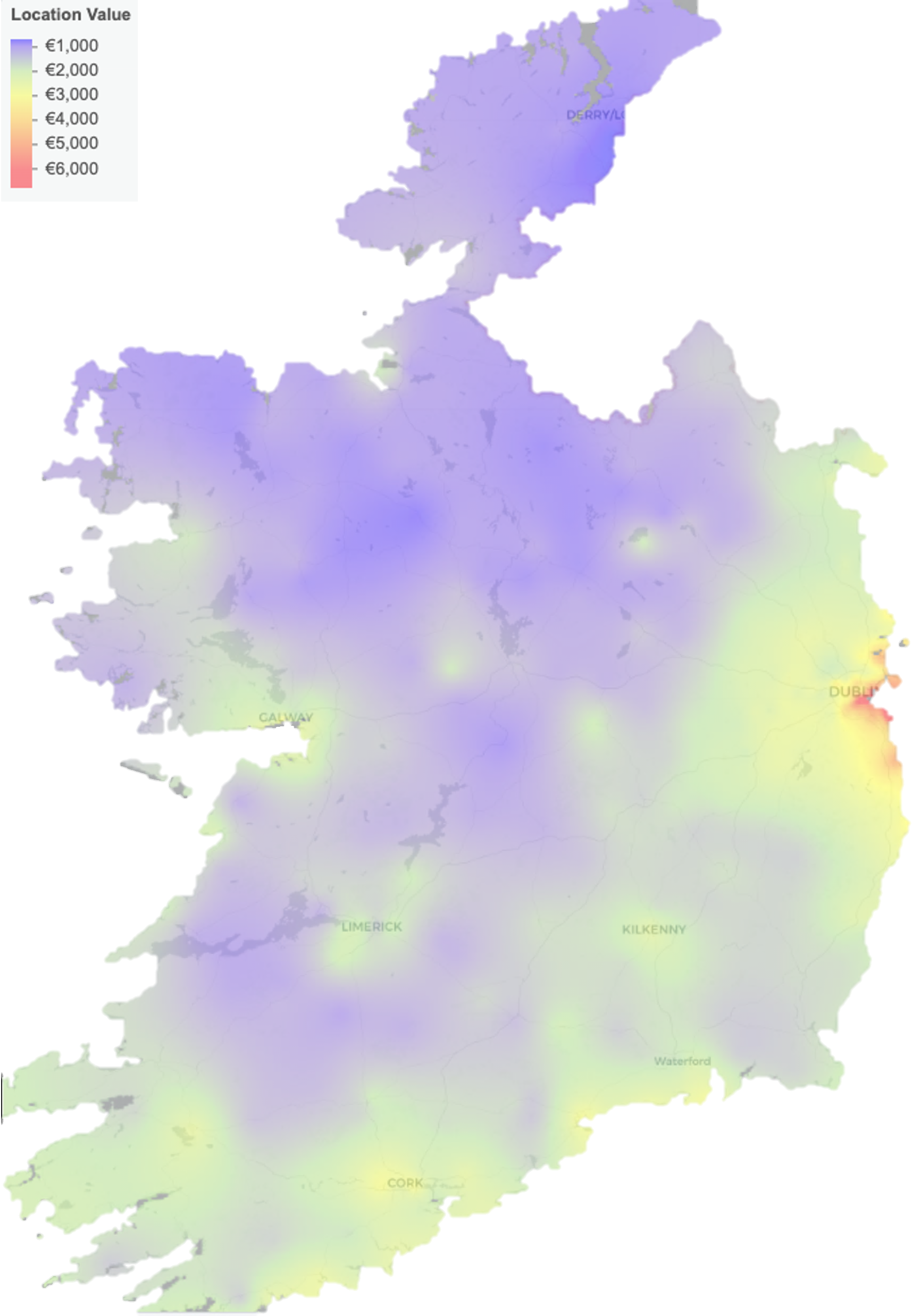}
    \caption{\small  Map of the predicted price per $m^2$ with the GP and MRF spatial surface}
    \label{fig:g2_GP_MRF_map_clipped}
\end{figure}
\end{center}

In Figure \ref{fig:g2_GP_MRF_map_clipped}, a proxy location value map is plotted on a price per $m^2$ scale. By using the GP effects from Figure \ref{fig:g2_GP_map_clipped} adjusted for Eircode regions using the MRF surface from Figure \ref{fig:g2_MRF_map_lbls}, this map represents the property price per $m^2$ at each location before property features are accounted for. The structure of the GP surface is apparent, with higher value ``hot-spots'' surrounding cities and coastal areas in the west, south and east. Lower value locations are apparent in the midlands and north, aligning with the median values of Eircode regions from Figure \ref{fig:map_median_ecode}.

\newpage

\section{Discussion \& Conclusion}\label{sec:discussion}

The development of automated valuation models has accelerated in recent years, taking advantage of highly granular data and adopting machine learning techniques. Issues such as the varying density of observations pose a challenge to traditional hedonic models, which lack spatial complexity, and more advanced machine learning models, which often require a large amount of training data.  In this article, the benefits of accounting for the varying density of observations using geospatial techniques are investigated. Using data from 29,458 property listings across Ireland, we develop a flexible spatiotemporal GAM and compare it against the benchmark hedonic model and RF. The GAM includes a GP surface, which can borrow information from densely populated towns and cities to estimate values in rural areas with a low observation count. Modelling Irish Eircode regions as a MRF facilitates spatial smoothing through macro-scale adjustments; thus, this spatial fusion accounts for effects at varying spatial scales. 

While the RF did outperform the S-GAM across common metrics at a national scale, this article demonstrates that national model metrics do not highlight true model performance. The RF lacks accuracy in sparsely populated rural areas of Ireland, with an $R^2$ of 0.52, while the S-GAM has an $R^2$ of 0.78 in such areas. The S-GAM additionally outperforms the RF when tested on observations in Irish towns, and achieves similar metrics to the RF when tested on observations in Irish cities. 

This analysis strongly relies on the rich information of the main Irish property listing websites, which allows for the extraction of property characteristics using text mining. The benefits of using a statistical model with this rich data come with the interpretability of each component in detail. Modelling across submarkets has provided a useful method of comparing model accuracy and has allowed for the comparison of property characteristics across submarkets. While these interpretations are specific to Ireland, such an approach could be followed in any property market with underlying submarkets or a varying density of observations. Use of the low-rank spatial spline was partly driven by the computational burden of fitting a spatial process across the entire dataset. Future work could explore higher-resolution spatial processes that better capture fine-scale spatial patterns by using GP approximations that more fully leverage all available data, such as the nearest neighbour GP \parencite{Datta02042016, paci2017analysis} or hierarchical covariance approximation \parencite{dearmon2024local}.

This study fills a gap in the literature by moving beyond Irish cities to consider the Irish property market as a whole, using highly detailed property listing data to develop a flexible spatiotemporal technique for property valuation. Our findings are of interest to decision-making bodies in Ireland and could be used to improve the current hedonic models used for property index calculation by the CSO. Irish property tax calculations and site value estimations are based on antiquated and coarse techniques, such as the homeowner's valuation or comparison to the sale price of neighbouring properties, which may not be correctly adjusted for differences in property location, characteristics or the time of sale.  Improvements in data quality could advance real estate research in Ireland, for example, the use of listing websites with rich information fields, rather than relying on the realtor's text description. The availability of such property listings in a suitable format would not only improve the quality of researchers but holds the potential to benefit governments, individuals and institutions alike.

\newpage

\section*{Appendix}\label{Appendix}

\begin{table}[h!]
\footnotesize
\centering
\caption{\label{tab:tab:g2_coefficients}Coefficient estimates and 95\% confidence intervals for the linear terms of the S-GAM}
\centering
\begin{tabular}[t]{>{\raggedright\arraybackslash}p{6cm}>{\raggedright\arraybackslash}p{3cm}>{\raggedright\arraybackslash}p{4cm}}
\toprule
Variable & Estimate & 95\% Confidence Interval\\
\midrule
\hspace{1em}\textit{Property Type} &  \vphantom{1} & \\
\addlinespace
Attic Conversion & 0.99 & {}[0.97, 1.00]\\
\addlinespace
Garden & 1.03 & {}[1.02, 1.03]\\
\addlinespace
Cul-de-sac & 1.00 & {}[1.00, 1.01]\\
\addlinespace
Garage & 1.03 & {}[1.02, 1.03]\\
\addlinespace
Renovated Property & 1.04 & {}[1.03, 1.04]\\
\addlinespace
Period Property & 1.11 & {}[1.10, 1.13]\\
\addlinespace
South Facing Property & 1.02 & {}[1.01, 1.02]\\
\addlinespace
Ground Floor Apartment & 0.99 & {}[0.98, 1.01]\\
\addlinespace
Second Floor Apartment & 1.00 & {}[0.98, 1.02]\\
\addlinespace
Penthouse Apartment & 1.15 & {}[1.12, 1.19]\\
\addlinespace
New Property & 1.01 & {}[1.01, 1.02]\\
\addlinespace
\hspace{1em}\textit{Property Type} &  & \\
\addlinespace
Apartment & 0.92 & {}[0.91, 0.94]\\
\addlinespace
Detached & 1.24 & {}[1.22, 1.26]\\
\addlinespace
Duplex & 0.89 & {}[0.87, 0.91]\\
\addlinespace
End-of-terrace & 1.01 & {}[1.00, 1.02]\\
\addlinespace
Semi-detached & 1.09 & {}[1.08, 1.10]\\
\addlinespace
Terraced & 0.96 & {}[0.95, 0.97]\\
\addlinespace
Townhouse & 0.93 & {}[0.92, 0.94]\\
\addlinespace
\hspace{1em}\textit{BER Values} &  & \\
\addlinespace
A1 & 1.15 & {}[1.06, 1.24]\\
\addlinespace
A2 & 1.16 & {}[1.14, 1.18]\\
\addlinespace
A3 & 1.12 & {}[1.10, 1.14]\\
\addlinespace
B1 & 1.08 & {}[1.05, 1.11]\\
\addlinespace
B2 & 1.05 & {}[1.03, 1.06]\\
\addlinespace
B3 & 1.04 & {}[1.03, 1.05]\\
\addlinespace
C1 & 1.00 & {}[1.00, 1.01]\\
\addlinespace
C2 & 0.99 & {}[0.99, 1.00]\\
\addlinespace
C3 & 0.98 & {}[0.97, 0.99]\\
\addlinespace
D1 & 0.96 & {}[0.95, 0.97]\\
\addlinespace
D2 & 0.95 & {}[0.94, 0.96]\\
\addlinespace
E1 & 0.93 & {}[0.92, 0.94]\\
\addlinespace
E2 & 0.92 & {}[0.91, 0.93]\\
\addlinespace
F & 0.88 & {}[0.87, 0.90]\\
\addlinespace
G & 0.85 & {}[0.83, 0.86]\\
\bottomrule
\end{tabular}
\end{table}

\newpage
\clearpage

\addcontentsline{toc}{chapter}{Bibliography}

\printbibliography
\end{document}